\documentclass[preprint,12pt,authoryear]{elsarticle}

\usepackage{amssymb}
\usepackage{amsmath}

\usepackage{lineno}

\modulolinenumbers[1]
\usepackage{hyperref}

\usepackage{bm}
\usepackage[margin=2cm]{geometry}
\usepackage{accents}
\usepackage{color}
\usepackage{xcolor}
\usepackage{soul}
\usepackage{multirow}
\usepackage{mathtools}

\journal{International Journal of Plasticity}

\newcommand{\ubar}[1]{\underaccent{\bar}{\bm{#1}}}
\newcommand{\tend}[1]{\hbox{\oalign{$\bm{#1}$\crcr\hidewidth$\scriptscriptstyle\bm{\sim}$\hidewidth}}}
\newcommand{\tene}[1]{\hbox{\oalign{$\bm{#1}$\crcr\hidewidth$\scriptscriptstyle\bm{\simeq}$\hidewidth}}}
\newcommand{\tenq}[1]{\hbox{\oalign{$\bm{#1}$\crcr\hidewidth$\scriptscriptstyle\bm{\approx}$\hidewidth}}}
\newcommand{\pvec}[2]{\hbox{\oalign{$\accentset{\times}{\underaccent{\bar}{\bm{#1}}}^{#2}$\crcr\hidewidth}}}
\newcommand{\pvecdot}[2]{\hbox{\oalign{$\accentset{\times}{\underaccent{\bar}{\dot{\bm{#1}}}}^{#2}$\crcr\hidewidth}}}

\newcommand{\mathcolorbox}[2]{\colorbox{#1}{$\displaystyle #2$}}
\newcommand{\ol}[1]{\overline{#1}}

\usepackage[english]{babel}
\addto\captionsenglish{}

\begin{document}

\begin{frontmatter}



\title{A multi-physics model for the evolution of grain microstructure} 

\author[fzj]{I.T. Tandogan}

\author[fzj]{M. Budnitzki\texorpdfstring{\corref{mycorrespondingauthor}}{}}
\cortext[mycorrespondingauthor]{Corresponding author}
\ead{m.budnitzki@fz-juelich.de}

\author[fzj,rwth]{S. Sandfeld}

\affiliation[fzj]{organization={Institute for Advanced Simulations – Materials Data Science and Informatics (IAS‑9), Forschungszentrum Jülich GmbH},
	addressline={},
	city={Jülich},
	postcode={52425},
	state={},
	country={Germany}}

\affiliation[rwth]{organization={Chair of Materials Data Science and Materials Informatics, Faculty 5 – Georesources and Materials Engineering, RWTH Aachen University},
	addressline={},
	city={Aachen},
	postcode={52056},
	state={},
	country={Germany}}

\begin{abstract}
When a metal is loaded mechanically at high temperatures, i.e. above 300 $^o$C, its grain microstructure evolves due to multiple physical mechanisms. Two of which are the curvature-driven migration of the grain boundaries due to increased mobility, and the formation of subgrains due to severe plastic deformation. Similar phenomena are observed during heat treatment subsequent to severe plastic deformation. Grain boundary migration and plastic deformation simultaneously change the lattice orientation at any given material point, which is challenging to simulate consistently. The majority of existing simulation approaches tackle this problem by applying separate, specialized models for mechanical deformation and grain boundary migration sequentially. Significant progress was made recognizing that the Cosserat continuum represents an ideal framework for the coupling between different mechanisms causing lattice reorientation, since rotations are native degrees of freedom in this setting. 

In this work we propose and implement a multi-physics model, which couples Cosserat crystal plasticity to Henry-Mellenthin-Plapp (HMP) type orientation phase-field in a single thermodynamically consistent framework for microstructure evolution. Compared to models based on the Kobayashi-Warren-Carter (KWC) phase-field, the HMP formulation removes the nonphysical term linear in the gradient of orientation from the free energy density, thus eliminating long-range interactions between grain boundaries. Further, HMP orientation phase field can handle inclination-dependent grain boundary energies.
We evaluate the model’s predictions and numerical performance within a two-dimensional finite element framework, and compare it to a previously published results based on KWC phase-field coupled with Cosserat mechanics.
\end{abstract}


\begin{highlights}
\item A unified multi-physics model for the evolution of grain microstructure is proposed
\item Orientation phase field and Cosserat crystal plasticity models are strongly coupled
\item Lattice reorientation by both deformation and grain boundary migration is possible
\item Grain boundary migration is driven by curvature and stored dislocation energy
\item Non-physical long range interactions of grain boundaries are removed
\item Read-Shockley type and inclination-dependent grain boundary energies are possible
\end{highlights}


\end{frontmatter}


\section{Introduction}

Forming of metallic alloys involves high stress deformation at elevated temperatures and heat treatment subsequent to severe plastic deformation, both of which extensively alter the granular microstructure and consequently the macroscopic material properties. Evolution of the polycrystalline structure comprises of grains of different sizes, shapes and orientations, and the corresponding grain boundary network can occur through multiple physical mechanisms including viscoplastic deformation, static and dynamic recrystallization, recovery, grain nucleation and growth \citep{rollett2017recrystallization}. Viscoplastic deformation changes the orientation of the crystal lattice, which in severe cases can lead to the formation of localized slip and kink bands \citep{asaro1977strain} and the accompanied fragmentation of grains into sub-grains \citep{sedlavcek2002subgrain}. Inside the grains, dislocations build up, often non-homogeneously, due to localized deformation and random trapping \citep{ashby1970deformation}. The resulting increase of energy is a driving force for the motion of grain boundaries, the mobility of which is related to misorientation \citep{sutton1995interfaces,gottstein2009grain}. Grain boundary motion is possible via coordinated shuffling of atoms, which also  reorients the crystal lattice. During recrystallization, grains tend to nucleate at locations of high misorientation, sites with high dislocation density and second-phase particles, which then grow into the other grains reducing the total energy.

These processes simultaneously change the lattice orientation at any given material point, which is challenging to simulate consistently. At atomistic length scale, molecular dynamics simulations were able to capture the complex physics of grain boundary deformation and motion \citep{cahn2006coupling,upmanyu2006simultaneous,mishin2010atomistic}. While they provide interesting insights, it is not feasible to apply them to the time and length scale of polycrystal evolution at mesoscale. Most of the mesoscale models treat the evolution mechanisms separately, covering either deformation or evolution of the grain structure. Classical crystal plasticity theories for bulk plasticity with fixed grain boundaries have been improved rigorously, and can accurately predict viscoplastic deformation in engineering problems \citep{roters2010overview}, while generalized non-local theories make it possible to incorporate physical mechanisms such as size effects, strain localization and grain boundary dislocation interaction \citep{forest2000cosserat,gurtin2000plasticity,mayeur2011dislocation,yalccinkaya2021misorientation}. The kinetics of grain boundary motion have been studied with a number of methods such as vertex techniques (cf. \cite{soares1985computer,gill1996variational}), cellular automata (e.g. \cite{marx1999simulation,raabe2002cellular}), Monte-Carlo \citep{srolovitz1986grain,rollett1992microstructural}, level-set methods \citep{chen1995novel,bernacki2008level} and phase field models, where the last two have the advantage of tracking moving interface implicitly. In particular, phase-field approaches for grain boundary migration are used extensively, which are divided mainly into multi-phase field models (see \cite{steinbach1996phase,steinbach1999generalized,fan1997computer}) and orientation phase field models (cf. \cite{kobayashi2000continuum,warren2003extending,henry2012orientation}). 

For the coupled evolution problem, the above mentioned independent methods have been combined mainly by employing them sequentially or through a phenomenological coupling. Usually, information is passed back and forth between a crystal plasticity (CP) model for deforming the structure and a model for grain boundary (GB) migration for evolving it, possibly with an intermediate step for the nucleation of grains based on stochastic methods or a trigger criteria such as misorientation or stored energy. \cite{mellbin2017extended} used finite strain CP and an extended vertex model successively for dynamic recrystallization and grain growth. Similarly, cellular automata (CA) approach was combined with CP to model static \citep{raabe2000coupling} and dynamic recrystallization \citep{popova2015coupled,LI2016154} where Li et al. included mechanical feedback from CA. \cite{bernacki2011level} and \cite{hallberg2013modified} employed a level-set framework combined with CP, while \cite{blesgen2017variational} used higher order Cosserat CP in a staggered scheme. In a series of contributions, \cite{takaki2008multi,takaki2009multi,takaki2010static,takaki2014multiscale} developed models coupling multi-phase field (MPF) with various deformation theories such as CP, strain gradient CP, Kocks-Mecking and J2 for static and dynamic recrystallization problems. \cite{chen2015integrated} and \cite{zhao2016integrated} efficiently simulated three dimensional static and dynamic recrystallization in polycrystals by using spectral Fast-Fourier-Transform (FFT) based MPF and CP solvers. \cite{abrivard2012aphase} and \cite{luan2020combining} coupled KWC type orientation phase field extended with a stored energy term to finite element CP. Unlike MPF, KWC has the advantage of representing lattice orientation as a degree of freedom. Although sequential combinations of different methods have been used successfully to model microstructure evolution along with deformation, several authors have sought a unified thermodynamically consistent field theory that strongly couples the evolution of mechanical and kinetic variables. \cite{cahn2006coupling}, \cite{frolov2012thermodynamics}, \cite{berbenni2013micromechanics} and \cite{basak2015simultaneous,basak2015three} have developed a class of sharp interface continuum models where shear loading can stimulate grain boundary motion. However, they are not tuned for the mesoscale, where interaction of bulk plasticity and grain boundaries is important. \cite{admal2018unified} and \citep{he2021polycrystal} have proposed a model that considers bulk and grain boundary plasticity in unison. A large deformation strain gradient CP is formulated with a non-standard free energy density containing contributions from the full dislocation density tensor inspired by the KWC type orientation phase field. Hence, it is capable of predicting both shear-induced and curvature driven GB motion, grain sliding, rotation and subgrain nucleation, where the GB motion is accommodated by plastic slip process. Orientation phase field models are well suited for a strong coupling with crystal plasticity as the lattice orientation is a continuous field in bulk and interface. \cite{ask2018cosserat,ask2018bcosserat,ask2019cosserat,ask2020microstructure} have combined the modified KWC orientation phase field of \cite{abrivard2012aphase} with Cosserat-type crystal plasticity with independent micro-rotation degrees of freedom, recognizing it as a natural framework for the coupling. Lattice orientation can evolve simultaneously due to viscoplastic deformation and GB migration, where the latter is driven by capillary forces as well as accumulated dislocation densities. The coupling of Cosserat continuum with orientation phase field allows heterogeneous reorientation and subgrain nucleation in the bulk. Recently, \cite{ghiglione2024cosserat} showed through the torsion of a single crystal rod that the model can predict spontaneous grain nucleation in the presence of a lattice orientation gradient.

In the present work, a model inspired by \cite{ask2018cosserat} is proposed where the Henry-Mellenthin-Plapp (HMP) type orientation phase field \citep{henry2012orientation,staublin2022phase} is strongly coupled with a Cosserat crystal plasticity (CCP) framework. The proposed model offers several improvements. By construction, the HMP model removes non-physical long range interactions between the grain boundaries that exist in the KWC model due to the $|\nabla\theta|$ term in the free energy. The HMP model achieves localized grain boundaries using a singular coupling function. This coupling function can be modified in order to obtain the desired misorientation dependence of grain boundary energy without changing the variational form of the model, for example inclination dependent energy. Moreover, the energy contribution of statistically stored dislocations to the free energy density, which was added by \cite{abrivard2012aphase} into KWC to drive GB motion, is modified. Originally this term altered the equilibrium state of the order parameter in the bulk, which is undesirable in a multi-phase context. We propose a form that allows similar GB kinetics without changing the phase field.

This paper is structured as follows. Section \ref{sec:model} summarizes the general coupled orientation phase field and Cosserat CP framework, presents the governing equations and evolution of history variables in the proposed coupled model, and ends by comparing it with the previous KWC+CCP framework by \cite{ask2018cosserat}. Section \ref{sec:numex} explores the fundamental mechanisms of the coupled model, establishing the model parameters with simple numerical examples, namely, equilibrium profiles, shear loading and dislocation driven GB migration in a periodic bicrystal structure, as well as the triple junction test for curvature driven GB motion. Finally, the paper is concluded with a summary and outlook in Section \ref{sec:conc}.

\section{Model}\label{sec:model}
This section presents the model framework, describing the kinematics, balance laws and the formulation of the constitutive laws within a small deformation setting. As a starting point Sections \ref{ssec:cosserat}-\ref{ssec:consev} give an overview of the coupled Cosserat-phase field model developed by \cite{ask2018cosserat} which was later modified slightly in \cite{ask2018bcosserat}, where Cosserat crystal plasticity was coupled with KWC type orientation phase field \citep{kobayashi2000continuum,warren2003extending}. Section \ref{ssec:freeen} proposes the Cosserat-HMP phase field model introducing a new free energy and some modifications, where we use a new phase field model for coupling \citep{henry2012orientation,staublin2022phase}. Section \ref{ssec:compkwcccp} highlights the differences of the proposed model with the Cosserat-KWC phase field model of \cite{ask2018cosserat,ask2018bcosserat}.

\subsection*{Notation}

In the following, vectors $a_i$ are denoted by $\ubar{a}$, 2nd order tensors $A_{ij}$ by $\tend{A}$, 3rd order Levi-Civita permutation tensor $\epsilon_{ijk}$ by $\tene{\epsilon}$, 4th order tensors $C_{ijkl}$ by $\tenq{C}$. Gradient is denoted by $\nabla(.)$, divergence by $\nabla\cdot(.)$, trace by $tr(.)$, transpose by $(.)^T$, dot product by $(.)\cdot(.)$, double contraction by $(.):(.)$, and tensor product by $(.)\otimes(.)$. The transformation between pseudo-vector $\pvec{a}{}$ and skew-symmetric tensor $\tend{A}^{\text{skew}}$ is given by,

\begin{equation}
	\pvec{a}{}=-\frac{1}{2}\tene{\epsilon}:\tend{A} \quad\text{and}\quad \tend{A}=-\tene{\epsilon}\cdot\pvec{a}{}.
\end{equation} 

\subsection{Cosserat continuum coupled with orientation phase field}\label{ssec:cosserat}

The Cosserat continuum introduces a microrotation tensor $\tend{R}$ which relates the current state of a triad of orthonormal directors to the initial state at each material point, independently from the displacements. In small deformation setting, it is given by,

\begin{equation}
	\tend{R}=\tend{I}-\tene{\epsilon}\cdot\ubar{\Theta}
\end{equation}
where $\tend{I}$ is the identity tensor and $\ubar{\Theta}$ is the microrotation pseudo-vector. The displacement vector $\ubar{u}$ together with $\ubar{\Theta}$ describes the motion of a material point with six degrees of freedom. The objective deformation measures are given by \citep{eringen1976polar,forest1997cosserat},

\begin{equation}
	\tend{e}=\nabla\ubar{u}+\tene{\epsilon}\cdot\ubar{\Theta}, \quad\quad \kappa=\nabla\ubar{\Theta},
\end{equation}
which are the deformation tensor and curvature tensor, respectively. The small deformation setting allows an additive decomposition of the strain into elastic and plastic contributions

\begin{equation}
	\tend{e} = \tend{e}^e + \tend{e}^p,
\end{equation}
where plastic curvature is not considered for simplicity. The rate of the deformation tensor can be expressed as

\begin{equation}
	\dot{\tend{e}} =\dot{\tend{\varepsilon}}+\tend{\omega}+\tene{\epsilon}\cdot\dot{\ubar{\Theta}},
\end{equation}
where,

\begin{equation}
	\dot{\tend{\varepsilon}} = \dfrac{1}{2}\left[\nabla\dot{\ubar{u}}+\left(\nabla\dot{\ubar{u}}\right)^T\right] \quad\text{and}\quad \tend{\omega} = \dfrac{1}{2}\left[\nabla\dot{\ubar{u}}-\left(\nabla\dot{\ubar{u}}\right)^T\right]
\end{equation}
are the strain rate and spin tensors, respectively. The skew-symmetric part of $\dot{\tend{e}}$ can be represented by the pseudo-vector

\begin{equation}
	\pvecdot{e}{} = \pvec{\omega}{}-\dot{\ubar{\Theta}},
\end{equation}
which relates the rotation of the material to the Cosserat microrotation. Using the elastic-plastic decomposition and defining plastic and elastic spin pseudo-vectors $\pvec{\omega}{p}:=\pvecdot{e}{p}$ and $\pvec{\omega}{e}:=\pvec{\omega}{}-\pvec{\omega}{p}$, respectively, we can write

\begin{equation}\label{eqn:rotlink}
	\pvecdot{e}{e} = \pvec{\omega}{e}-\dot{\ubar{\Theta}}.
\end{equation}
Equation \eqref{eqn:rotlink} expresses a link between the rotation rate of the crystal lattice and the Cosserat microrotation rate, which are distinct at this point. For our purposes they should be identical, which can be enforced by the internal constraint,

\begin{equation}\label{eqn:cosconstraint}
	\pvec{e}{e}\equiv 0,
\end{equation}
which is possible to implement on the constitutive level with a penalty parameter \citep{forest2000cosserat,mayeur2011dislocation,blesgen2014deformation}.

In order to include grain boundary migration, the Cosserat theory is enhanced with an orientation phase field model. The latter defines the microstructure with the order parameter $\eta$, lattice orientation $\ubar{\Theta}$ and their gradients $\nabla\eta$ and $\nabla\ubar{\Theta}$. The order parameter $\eta\in [0,1]$ is a coarse-grained measure of crystalline order that takes the value $\eta=1$ in the bulk of the grain and $\eta<1$ in diffuse grain boundaries. The coupling of Cosserat continuum with the phase field is explained next.

\subsection{Balance laws}\label{ssec:balance}

By using the microforce formalism described by Gurtin, the phase field equations can be incorporated into equations of mechanics \citep{gurtin1996generalized,ammar2009finite}. The reader is referred to \cite{ask2018cosserat} for the detailed derivation steps. The principle of virtual power is applied using following set of virtual rates,

\begin{equation}
	\mathcal{V} = \left\{\dot{\eta}^\star,\nabla\dot{\eta}^\star,\dot{\ubar{u}}^\star,\nabla\dot{\ubar{u}}^\star,\dot{\ubar{\Theta}}^\star,\nabla\dot{\ubar{\Theta}}^\star\right\},
\end{equation} 
and the resulting balance equations and boundary conditions are given by,

\begin{alignat}{3}
	&\nabla\cdot\ubar{\xi}_\eta + \pi_\eta +\pi_\eta^{\text{ext}}=0 \quad\quad\quad &&\text{in}\;\;\Omega,&&\label{eqn:balgen}\\
	&\nabla\cdot\tend{\sigma}+\ubar{f}^{\text{ext}}=\ubar{0} \quad\quad\quad &&\text{in}\;\;\Omega,&&\label{eqn:ballin}\\
	&\nabla\cdot\tend{m}+2\pvec{\sigma}{}+\ubar{c}^{\text{ext}}=\ubar{0} \quad\quad\quad &&\text{in}\;\;\Omega,&&\label{eqn:balang}\\
	&\ubar{\xi}_\eta\cdot\ubar{n}=\pi_\eta^\text{c} \quad\quad\quad &&\text{on}\;\;\partial\Omega,&&\\
	&\tend{\sigma}\cdot\ubar{n}=\ubar{f}^\text{c} \quad\quad\quad &&\text{on}\;\;\partial\Omega,&&\\
	&\tend{m}\cdot\ubar{n}=\ubar{c}^\text{c} \quad\quad\quad &&\text{on}\;\;\partial\Omega.&&
\end{alignat}
The generalized stresses $\pi_\eta$ and $\ubar{\xi}_\eta$ correspond to Gurtin's microforces and are conjugate to the order parameter $\eta$ and its gradient $\nabla\eta$. The stress $\tend{\sigma}$ which contains a skew-symmetric part as expected in Cosserat continuum is conjugate to the Cosserat deformation $\tend{e}$. The couple-stress $\tend{m}$ is conjugate to curvature $\tend{\kappa}$. Equations \eqref{eqn:balgen}, \eqref{eqn:ballin} and \eqref{eqn:balang} represent balance of generalized stresses, balance of linear momentum and balance of angular momentum, respectively. Superscript $(.)^\text{ext}$ denotes external body forces and couple forces, while $(.)^\text{c}$ denotes contact forces and couples. The outward normal to the surface $\partial\Omega$ of volume $\Omega$ is $\ubar{n}$.

\subsection{Constitutive and evolution equations}\label{ssec:consev}

The formulation of constitutive and evolution equations are slightly different between \cite{ask2018cosserat} and \cite{ask2018bcosserat} in their treatment of dissipation potential and an eigendeformation term. Firstly, in the second publication the dissipative contribution to the stress $\tend{\sigma}$ is dropped. Previously, this resulted in a separate evolution term of $\Theta$ similar to the phase field model, which becomes redundant in the coupled model. Secondly, the eigendeformation is considered as a part of plastic deformation rather than as a separate strain, whose function is explained in Section \ref{ssec:freeen}. This work follows the approach of \cite{ask2018bcosserat} as summarized below. 

We assume that the Helmholtz free energy density $\Psi$ is defined as,

\begin{equation}
	\rho\Psi\coloneqq\psi\left(\eta,\nabla\eta,\tend{e}^e,\tend{\kappa},r^\alpha\right)
\end{equation}
where $r^\alpha$ are internal variables related to plasticity. Applying the Clausius-Duhem inequality results in,

\begin{equation}
	-\left[\pi_\eta+\dfrac{\partial\psi}{\partial\eta}\right]\dot{\eta} + \left[\ubar{\xi}_\eta-\dfrac{\partial\psi}{\partial\nabla\eta}\right]\cdot\nabla\dot{\eta} + \left[\tend{\sigma}-\dfrac{\partial\psi}{\partial\tend{e}^e}\right]:\dot{\tend{e}}^e + \left[\tend{m}-\dfrac{\partial\psi}{\partial\tend{\kappa}}\right]:\dot{\tend{\kappa}} +\tend{\sigma}:\dot{\tend{e}}^p-\sum_\alpha\dfrac{\partial\psi}{\partial r^\alpha}\dot{r}^\alpha \ge 0 .
\end{equation}

In order to recover the relaxation behavior of the phase field model, the stress $\pi_\eta$ is assumed to contain energetic and dissipative contributions such that

\begin{equation}
	\pi_\eta = \pi_\eta^\text{eq} + \pi_\eta^\text{neq}.
\end{equation}
Then, the constitutive relations are given by

\begin{equation}\label{eqn:const}
	\pi_\eta^{eq}=-\dfrac{\partial\psi}{\partial\eta},\qquad\ubar{\xi}_\eta=\dfrac{\partial\psi}{\partial\nabla\eta},\qquad\tend{\sigma}=\dfrac{\partial\psi}{\partial\tend{e}^e},\qquad\tend{m}=\dfrac{\partial\psi}{\partial\tend{\kappa}}.
\end{equation}
The dissipation potential $\Omega$ is assumed to contain following contributions

\begin{equation}
	\Omega=\Omega^p(\tend{\sigma}) + \Omega^\alpha(R^\alpha)+\Omega^\eta(\pi_\eta^{neq}),
\end{equation}
where

\begin{equation}
	R^\alpha=\dfrac{\partial\psi}{\partial r^\alpha}
\end{equation}
are thermodynamic forces related to $r^\alpha$. The plastic flow, hardening rules and evolution equations are found from $\Omega$ as

\begin{equation}\label{eqn:evodot}
	\dot{\tend{e}}^p=\dfrac{\partial\Omega^p}{\partial\tend{\sigma}},\qquad \dot{r}^\alpha=-\dfrac{\partial\Omega^\alpha}{\partial R^\alpha},\qquad \dot{\eta}=-\dfrac{\partial\Omega^\eta}{\partial\pi_\eta^{neq}}.
\end{equation}

\subsection{Coupled model}\label{ssec:freeen}

For the coupled problem, a general form of free energy functional including phase field variables and deformation measures is considered:
\begin{align}\label{eqn:freeen3d}
	\begin{split}
		\psi\left(\eta,\nabla\eta,\tend{e}^e,\tend{\kappa},r^\alpha\right) &= f_0\left[\alpha V(\eta)+\frac{\nu^2}{2}|\nabla\eta|^2+\mu^2g(\eta)a^2(\ubar{n},\ubar{\Theta})||\tend{\kappa}||^2\right]\\
		&+\dfrac{1}{2}\tend{\varepsilon}^e:\tenq{E}^s:\tend{\varepsilon}^e+2\mu_c(\eta)\,\pvec{e}{e}\cdot\pvec{e}{e}+\psi_\rho(\eta,r^\alpha).
	\end{split}
\end{align}
The first line of \eqref{eqn:freeen3d} is a generalization of Henry-Mellenthin-Plapp \citep{henry2012orientation} free energy to three dimensions, which is multiplied with normalization parameter $f_0$ with units of Pa or J/m$^3$. The potential $V(\eta)$ penalizes the existence of grain boundaries, where $\eta<1$, and depending on whether it is a double-well or single-well potential, it is possible to have solid-liquid phases or solid phase only. For our purposes we choose the latter similar to \cite{staublin2022phase}. The second and third terms are grain boundary terms due to order parameter $\eta$ and a contribution due to lattice curvature $\tend{\kappa}$, penalizing gradients in $\eta$ and $\ubar{\Theta}$, respectively, where the coefficients $\nu$ and $\mu$ with unit m and non-dimensional $\alpha$ set the length scale of the diffuse grain boundary. Localized grain boundary solutions are made possible by the third term through $g(\eta)$, the singular coupling function, relating lattice orientation to the phase field, which tends to infinity as $\eta\rightarrow1$. The anisotropy coefficient $a(\ubar{n},\ubar{\Theta})$
can be used to introduce an inclination dependence of the grain boundary energy, where $\ubar{n}$ is the grain boundary normal. 

The second line of \eqref{eqn:freeen3d} contains the energy contribution due to symmetric and skew-symmetric elastic deformation, where $\tenq{E}^s$ is 4th order elasticity tensor and $\mu_c(\eta)$ is the Cosserat couple modulus which can possibly have different values depending on order parameter. If $\mu_c$ is sufficiently large, the skew-symmetric part of the deformation is penalized, which enforces the constraint \eqref{eqn:cosconstraint}; then the Cosserat microrotation follows the lattice orientation. The last term is the energy contribution due to accumulated dislocations given by

\begin{equation}
	\psi_\rho(\eta,r^\alpha)=\phi(\eta)\sum_{\alpha=1}^{N}\frac{\lambda}{2}\mu^er^{\alpha\,2},
\end{equation}
where N is the number of slip systems, $\lambda$ is a parameter close to 0.3 \citep{hirth1983theory}, $\mu^e$ is the shear modulus and $r^\alpha$ are internal variables associated to statistically stored dislocations (SSD). $\phi(\eta)$ is a coupling function depending on phase field, which allows SSDs to act as a driving force for migration of grain boundaries.

Restricting the problem to two dimensions allows considerable simplification of the equations. In 2D, microrotations reduce to $\ubar{\Theta}=\left[0\;0\;\theta\right]^T$, hence it follows that $\pvec{\omega}{}=\left[0\;0\;\omega\right]^T$ and $\pvec{e}{}=\left[0\;0\;\accentset{\times}{e}\right]^T$. Similarly, the couple-stress $\tend{m}$ reduces to $\ubar{m}_\theta=[m_{31}\;m_{32}\;m_{33}]^T$. Further assuming an isotropic grain boundary energy and constant $\mu_c$, equation \eqref{eqn:freeen3d} simplifies to

\begin{align}\label{eqn:freeen2d}
	\begin{split}
		\psi\left(\eta,\nabla\eta,\tend{e}^e,\nabla\theta,r^\alpha\right) &= f_0\left[\alpha V(\eta)+\frac{\nu^2}{2}|\nabla\eta|^2+\mu^2g(\eta)|\nabla\theta|^2\right]\\
		&+\dfrac{1}{2}\tend{\varepsilon}^e:\tenq{E}^s:\tend{\varepsilon}^e+2\mu_c\,\accentset{\times}{e}^e\,^2+\phi(\eta)\sum_{\alpha=1}^{N}\frac{\lambda}{2}\mu^er^{\alpha\,2}.
	\end{split}
\end{align}
Inserting the free energy \eqref{eqn:freeen2d} into  \eqref{eqn:const}, we obtain the state laws

\begin{align}
	\pi_\eta^{eq}=& -f_0\left[\alpha V_{,\eta}+\mu^2g_{,\eta}|\nabla\theta|^2\right]-\phi_{,\eta}\sum_{\alpha=1}^{N}\frac{\lambda}{2}\mu^er^{\alpha\,2},\label{eqn:pieq}\\
	\ubar{\xi}_\eta=& f_0\left[\nu^2\nabla\eta\right],\label{eqn:xieta}\\
	\ubar{m}_\theta=& f_0\left[2\mu^2g(\eta)\nabla\theta\right],\label{eqn:mtheta}\\
	\tend{\sigma}=& \tenq{E}^s:\tend{\varepsilon}^e - 2\mu_c\,\tene{\epsilon}\cdot\pvec{e}{e},\label{eqn:sigall}
\end{align}
where $(.)_{,\eta}$ denotes partial derivative w.r.t the order parameter $\eta$. The last term in \eqref{eqn:sigall} is the skew part of the stress tensor and it is equivalent to $\pvec{\sigma}{}=2\mu_c\,\pvec{e}{e}$ or $\accentset{\times}{\sigma}=2\mu_c\,\accentset{\times}{e}^e$ in 2D.

The plastic deformation mechanism is implemented through a Cosserat crystal plasticity framework. The plastic dissipation potential is chosen as,

\begin{equation}\label{eqn:plaspot}
	\Omega^p=\sum_{\alpha=1}^N\dfrac{K_v}{n+1}\left<\dfrac{|\tau^\alpha|-R^\alpha/\phi(\eta)}{K_v}\right>^{n+1}+\dfrac{1}{2}\tau_*^{-1}(\eta,\nabla\eta,\nabla\theta)\pvec{\sigma}{}\cdot\pvec{\sigma}{},
\end{equation}
where $\langle\cdot\rangle$ are Macaulay brackets, $K_v$ and $n$ are viscosity parameters, $\tau^\alpha$ is the resolved shear stress and $R^\alpha$ is the critically resolved shear stress of slip system $\alpha$. From the free energy functional we have,

\begin{equation}
	R^\alpha=\dfrac{\partial\psi}{\partial r^\alpha}=\lambda\phi(\eta)\mu^e r^\alpha=\lambda\phi(\eta)\mu^e b\sqrt{\sum_{\beta=1}^{N}h^{\alpha\beta}\rho^\beta},
\end{equation}
where internal variables $r^\alpha$ are related to the SSD densities $\rho^\alpha$ as

\begin{equation}
	r^\alpha=b\sqrt{\sum_{\beta=1}^{N}h^{\alpha\beta}\rho^\beta}.
\end{equation}
$b$ is the norm of the Burgers vector of the considered slip system family and $h^{\alpha\beta}$ is the interaction matrix. $\tau^\alpha$ is the projection of stress \tend{\sigma} on the slip system $\alpha$

\begin{equation}
	\tau^\alpha=\ubar{l}^\alpha\cdot\tend{\sigma}\cdot\ubar{n}^\alpha,
\end{equation}
where $\ubar{l}^\alpha$ is the slip direction and $\ubar{n}^\alpha$ is the slip plane normal. Since in the Cosserat framework the stress tensor is generally non-symmetric, the skew part contributes an additional term to the resolved shear stress which can considered to be a size dependent kinematic hardening \citep{sedlavcek2000non,forest2003plastic,forest2008some}. Substituting \eqref{eqn:plaspot} to \eqref{eqn:evodot}, we get the evolution of plastic deformation

\begin{equation}\label{eqn:epdot}
	\dot{\tend{e}}^p=\sum_{\alpha=1}^N\dot{\gamma}^\alpha\ubar{l}^\alpha\otimes\ubar{n}^\alpha+\dfrac{1}{2}\tau_*^{-1}(\eta,\nabla\eta,\nabla\theta)\text{skew}(\tend{\sigma})
\end{equation}
with

\begin{equation}
	\dot{\gamma}^\alpha=\left<\dfrac{|\tau^\alpha|-R^\alpha}{K_v}\right>^n \text{sign}\,\tau^\alpha
\end{equation}
as the slip rate according to the viscoplastic flow rule from \cite{cailletaud1992micromechanical}. While the first terms in \eqref{eqn:plaspot} and \eqref{eqn:epdot} are classical, the second terms are included to account for atomic reshuffling happening at the grain boundaries during migration which results in reorientation \citep{ask2018bcosserat}. Hence, it is supposed be active only in the grain boundary region which is possible if $\tau_*(\eta,\nabla\eta,\tend{\kappa})$ is formulated to be small inside grain boundary and large in bulk. It is a purely skew-symmetric contribution acting only on the plastic spin. We can write

\begin{equation}
	\pvec{e}{p}=\pvec{e}{\text{slip}}+\pvec{e}{*},
\end{equation}
where

\begin{alignat}{3}
	\pvecdot{e}{\text{slip}}&=\pvec{\omega}{p}-\pvec{\omega}{*},\qquad\qquad \tend{\omega}^p-&&\tend{\omega}^* =&&\,\text{skew}\left(\sum_{\alpha=1}^N\dot{\gamma}^\alpha\ubar{l}^\alpha\otimes\ubar{n}^\alpha\right),\\
	\pvecdot{e}{*}&=\pvec{\omega}{*}, &&\pvec{\omega}{*} =&&\tau_*^{-1}(\eta,\nabla\eta,\nabla\theta)\pvec{\sigma}{}.\label{eqn:wstar}
\end{alignat}
$\pvec{e}{*}$ is introduced to make sure that initial stresses in a polycrystal made of grains with different orientations are zero. Consider, for example elastic skew-symmetric deformation,

\begin{equation}\label{eqn:eeskew}
	\pvec{e}{e}=\accentset{\times}{skew}(\nabla u)-\pvec{e}{p}-\ubar{\Theta}=\accentset{\times}{skew}(\nabla u)-\pvec{e}{slip}-\pvec{e}{*}-\ubar{\Theta}.
\end{equation}
If Cosserat microrotation represents the lattice orientation and deformations are zero, clearly elastic deformation is non-zero, unless we set,

\begin{equation}
	\pvec{e}{*}(t=0)=\pvec{e}{p}(t=0)=-\ubar{\Theta}(t=0).
\end{equation}
This non-zero initial condition of plastic slip is adopted following \cite{admal2018unified} and \cite{ask2018bcosserat}. From \eqref{eqn:eeskew}, it is seen that a moving grain boundary results in $\pvec{e}{e}$ and $\pvec{\sigma}{}$ in region of moving front, which are relaxed by the evolution of $\pvec{e}{*}$ in \eqref{eqn:wstar}.

During plastic deformation, the SSD density evolves by multiplication and annihilation mechanisms following a modified Kocks-Mecking-Teodosiu law \citep{abrivard2012aphase,ask2018cosserat,ask2018bcosserat}, such that

\begin{equation}\label{eqn:rhodot}
	\dot{\rho}^\alpha=
	\begin{cases}
		\dfrac{1}{b}\left(K\sqrt{\sum _\beta\rho^\beta}-2d\rho^\alpha\right)|\dot{\gamma}| -\rho^\alpha C_DA(|\nabla\theta|)\dot{\eta} \quad&\text{if}\quad \dot{\eta}>0\vspace{1mm}\\
		\dfrac{1}{b}\left(K\sqrt{\sum _\beta\rho^\beta}-2d\rho^\alpha\right)|\dot{\gamma}| &\text{if}\quad \dot{\eta}\le 0
	\end{cases}
	,
\end{equation}
where the extra term accounts for the static recovery of dislocations in the wake of a sweeping grain boundary \citep{abrivard2012aphase,ask2018cosserat,bailey1962recrystallization}. $K$ is the mobility constant and $d$ is the critical annihilation distance between dislocations of opposite sign. For sufficiently high $C_D$, full recovery in the wake is achieved, and the function $A(|\nabla\theta|)=\tanh(C_A^2|\nabla\theta|^2)$ localizes this recovery to grain boundary region where $C_A$ has unit m. The evolution of the order parameter $\eta$ is governed by the quadratic dissipation potential

\begin{equation}
	\Omega^\eta = \frac{1}{2}\tau_\eta^{-1}\pi_\eta^{neq\;2}
\end{equation}
similar to \cite{gurtin1996generalized} and \cite{abrivard2012aphase} where $\tau_\eta(\eta,\nabla\eta,\nabla\theta,T)$ is a positive scalar function. Applying \eqref{eqn:evodot}, we get,

\begin{equation}
	\pi_\eta^{neq}=-\tau_\eta\dot{\eta}.\label{eqn:pineq}
\end{equation}
Inserting \eqref{eqn:pieq}-\eqref{eqn:mtheta} and \eqref{eqn:pineq} into the balance laws \eqref{eqn:balgen} and \eqref{eqn:balang}, while assuming zero body forces and couples forces, gives the governing equations for the order parameter $\eta$ and lattice orientation $\theta$:

\begin{align}
	\tau_\eta\dot{\eta}&=f_0\nu^2\nabla^2\eta-f_0\left[\alpha V_{,\eta}+\mu^2g_{,\eta}|\nabla\theta|^2\right]-\mathcolorbox{lightgray}{\phi_{,\eta}}\sum_{\alpha=1}^{N}\frac{\lambda}{2}\mu^er^{\alpha\,2}\label{eqn:hmpetadot}\\
	0&=f_0\nabla\cdot\left[\mu^2g(\eta)\nabla\theta\right]+2\mu_c\,\accentset{\times}{e}^e\label{eqn:hmpthetadot}
\end{align}
where differences from \cite{ask2018cosserat,ask2018bcosserat,ask2019cosserat} are highlighted to be discussed in Section \ref{ssec:compkwcccp}. \eqref{eqn:hmpthetadot} can be rewritten using \eqref{eqn:wstar} as,

\begin{equation}\label{eqn:tdot_hmpccp}
	-\tau_*(\eta,\nabla\eta,\nabla\theta)\accentset{\times}{\dot{e}}^*=f_0\nabla\cdot\left[\mu^2g(\eta)\nabla\theta\right]
\end{equation}
Setting $\mu_c=0$ recovers the HMP evolution equations (cf. \cite{henry2012orientation,staublin2022phase}) without $\theta$ relaxation and with the addition of an SSD term.

In the free energies \eqref{eqn:freeen3d} or \eqref{eqn:freeen2d}, the choice of potential $V(\eta)$ and singular coupling function $g(\eta)$ of the phase field model requires some care as described by \cite{henry2012orientation} and \cite{staublin2022phase}. Following conditions should be met:

\begin{enumerate}
	\item The value of the potential in the bulk of the grain is $V(1)=0$
	\item 1 is a minimum: $V'(1)=0$
	\item The potential should have only a single minimum: $V''(\eta)>0$ for all $\eta$.
	\item To enable a uniformly strained solid (small constant $\nabla\theta$ ) to be
	stable against the spontaneous formation of grain boundaries,
	we follow \cite{henry2012orientation}, and impose the constraint $9V''(1)>V'''(1)$
	when the coupling function $g(\eta)\approx(1-\eta)^{-2}$. This condition is
	irrelevant if we
	choose $g(\eta)\approx(1-\eta)^{-3}$ and thus this divergence of $g(\eta)$ stabilizes localized boundaries against uniform strain regardless. This condition should be reconsidered if a different $g(\eta)$ is used.
\end{enumerate}
We choose a simple second order single-well potential satisfying above conditions:

\begin{equation}
	V(\eta)=\frac{1}{2}(1-\eta)^2
\end{equation}
and adopt the singular coupling function in \cite{staublin2022phase},

\begin{equation}\label{eqn:fung}
	g(\eta)=\dfrac{7\eta^3-6\eta^4}{(1-\eta)^3}.
\end{equation}

\subsection{Comparison with KWC+CCP}\label{ssec:compkwcccp}
The main form of the governing equations proposed by \cite{ask2018bcosserat,ask2019cosserat} are as follows,
\begin{align}
	\tau_\eta\dot{\eta}&=f_0\nu^2\nabla^2\eta-f_0\left[V_{,\eta}+\mathcolorbox{lightgray}{sg^{*}_{,\eta}|\nabla\theta|}+\frac{\epsilon^2}{2}h^{*}_{,\eta}|\nabla\theta|^2\right]-\mathcolorbox{lightgray}{\hspace{1mm}}\sum_{\alpha=1}^{N}\frac{\lambda}{2}\mu^er^{\alpha\,2}\label{eqn:kwcetadot}\\
	0&=\frac{f_0}{2}\nabla\cdot\left[\mathcolorbox{lightgray}{sg^{*}(\eta)\dfrac{\nabla\theta}{|\nabla\theta|}}+\epsilon^2h^{*}(\eta)\nabla\theta\right]+2\mu_c\accentset{\times}{e}^e\label{eqn:kwcthetadot},
\end{align}
where the main differences compared to proposed HMP+CCP equations \eqref{eqn:hmpetadot} and \eqref{eqn:hmpthetadot} are highlighted. The biggest difference stems from usage of different orientation phase field models, that is the existence of the linear $|\nabla\theta|$ term in \eqref{eqn:kwcetadot} and the corresponding singular term in \eqref{eqn:kwcthetadot}. This term is necessary and responsible in KWC for the localization of grain boundaries as it is penalizing a solution with uniform gradient of orientation. However, as it is singular inside the grains, it requires regularization during the numerical solution \citep{kobayashi2000continuum,warren2003extending}. Moreover, the singular diffusion equation allows long-range interactions between grain boundaries, which results in a rotation rate for grains independent on the distance between grain boundaries \citep{kobayashi1999equations}. Hence, in the KWC model the only way to prevent unwanted grain rotations is to use a custom mobility function taking small values inside the grain \citep{warren2003extending}. On the other hand, the HMP model omits the linear $|\nabla\theta|$ term in free energy, and instead uses a singular coupling function $g(\eta)$ which approaches infinity inside grains. As a consequence, it has local interaction of grain boundaries and a constant mobility is sufficient to suppress grain rotations \citep{henry2012orientation}. 

Another difference is the way the atomic reshuffling process, i.e. the term $\tau_*(\eta,\nabla\eta,\nabla\theta)\accentset{\times}{\dot{e}}^*$ in \eqref{eqn:tdot_hmpccp}, is localized to grain boundaries. In the series of papers on the KWC+CCP model, \cite{ask2018cosserat,ask2019cosserat,ask2020microstructure} have used various forms such as,

\begin{equation}\label{eqn:locfunc}
	\tau_*=\hat{\tau}_*\tanh^{-1}(C_*^2|\nabla\eta|^2),\quad \tau_*=\hat{\tau}_*\tanh^{-1}(C_*^2|\nabla\theta|^2),\quad \tau_*=\hat{\tau}_*\left(1-\left[1-\dfrac{\mu_p}{\varepsilon}\right]\exp(-\beta_P\varepsilon|\nabla\theta|)\right)
\end{equation}
all of which are large in the bulk of grain and small at the grain boundary, with slightly different dynamics. In the proposed HMP+CCP model such a function already exists, i.e. the singular coupling function $g(\eta)$. Hence, we propose,

\begin{equation}\label{eqn:taustarg}
	\tau_*=\hat{\tau}_*g(\eta)
\end{equation}
Looking at equation \eqref{eqn:tdot_hmpccp}, using \eqref{eqn:taustarg} is a translation of the idea proposed by \cite{korbuly2017topological} to the coupled model, where they modify their mobility coefficient related to $\dot{\theta}$ in the same way in order to counterbalance the divergence of $g(\eta)$ multiplying $\nabla\theta$ in \eqref{eqn:hmpthetadot}. This modification significantly improves the convergence behavior of the model, and we explore its implications for the Cosserat mechanics.

Finally, a modification of the SSD energy term in \eqref{eqn:hmpetadot} or \eqref{eqn:kwcetadot} is proposed. In the latter, the multiplier function $\phi(\eta)$ of this term is simply $\eta$ and its derivative is 1. This has an important consequence on the equilibrium value of the order parameter inside the grains when dislocation density $\rho^\alpha$ is non-zero. We can easily calculate it from \eqref{eqn:hmpetadot} or \eqref{eqn:kwcetadot}. At equilibrium gradient terms and rate terms are zero, so we have,

\begin{equation}
	0=f_0\alpha(1-\eta^{eq})-\phi_{,\eta}(\eta^{eq})\psi_\rho(r^\alpha)
\end{equation}
or

\begin{equation}\label{eqn:etaeq}
	\eta^{eq}=1-\dfrac{\phi_{,\eta}(\eta^{eq})\psi_\rho(r^\alpha)}{f_0\alpha}.
\end{equation}
It is clear that when $\phi_{,\eta}=1$, the equilibrium value of the order parameter becomes less then 1, which is a change compared to the pure phase field model \citep{abrivard2012aphase,ask2018cosserat}. This is a significant problem for the formulation of HMP phase field since it relies on the singularity of $g(\eta)$ which is not singular if $\eta^{eq}\neq 1$. We believe this could also be problematic if more complex potentials $V(\eta)$ with temperature dependence and minima for different phases are considered, so a solution is proposed to circumvent this. In \eqref{eqn:etaeq}, it can be seen that if the derivative is taken to be zero inside the grain, i.e. $\phi_{,\eta}(\eta=1)=0$, then $\eta^{eq}$ remains as one independent of the SSD density. However, since this term is also the driving force for grain boundary migration due to dislocations, the choice of the function affects the grain boundary mobility. We have explored several polynomial forms and concluded that some of them can fit our purpose without modifying the dynamics considerably, which is discussed in the numerical examples later on. 

\section{Numerical Examples}\label{sec:numex}

In this section, the proposed model and its capabilities are tested through numerical examples simulated with the finite element method, and the results are discussed. The geometry shown in Fig. \ref{fig:bicrystal} is used to examine equilibrium profiles at the grain boundaries, elastic shear loading and grain boundary migration due to stored dislocation density. The classical triple-junction problem for testing phase field models, shown in Fig. \ref{fig:tripjunc}, is used to confirm our numerical implementation.

For these examples we consider the two dimensional form of the model governed by equations \eqref{eqn:hmpetadot}, \eqref{eqn:hmpthetadot} and \eqref{eqn:ballin}. Small deformation and plane strain conditions are assumed. 
We use isotropic grain boundaries for simplicity, and focus on the mechanisms made possible by Cosserat coupling. The simulated material is pure copper (Cu), and we do not pursue a rigorous fitting to experimental data at this stage; instead the model parameters are chosen to be in a reasonable range (see e.g. \cite{tschopp2015symmetric}). As presented in \ref{app:param}, the asymptotic analysis for HMP phase field can also be used for the coupled model to obtain equilibrium profiles and isotropic grain boundary energy for given misorientations. The parameters are calibrated to have 0.5 J/m$^2$ grain boundary energy at 15 degree misorientation. Unless otherwise mentioned, the model parameters given in Table \ref{tab:param} are used in the simulations.

The model has been implemented in the FEniCS 2019 open-source finite element library (\cite{logg2012automated}, \cite{alnaes2015fenics}). The details of the implementation including the weak forms, update of state variables and some numerical considerations are presented in \ref{app:fenics}. The system of equations is solved with a monolithic approach where, in two dimensions, each node has 4 degrees of freedom: order parameter $\eta$, Cosserat microrotation $\theta$ and 2 displacements $(u_1,u_2)$. A semi-implicit time discretization is used, and the resulting nonlinear system of equations is solved with the iterative Newton-Raphson algorithm.
See \ref{app:fenics} for details. Enforcing constraint \eqref{eqn:cosconstraint} allows us to identify the lattice orientation with Cosserat microrotation $\ubar{\Theta}$, which has only the third component in two-dimensions, the in-plane rotation $\theta$. It is measured with respect to the fixed global coordinate system given in Fig. \ref{fig:bicrystal}. The constitutive equations are applied in the local material frame. A feature of the coupled model is that the continuous and evolving lattice orientation field $\theta$ is used to rotate between global and the material coordinate system.

\subsection{Equilibrium profiles}\label{sec:equi}

\begin{figure}[ht]
	\centering
	\includegraphics[width=1\textwidth]{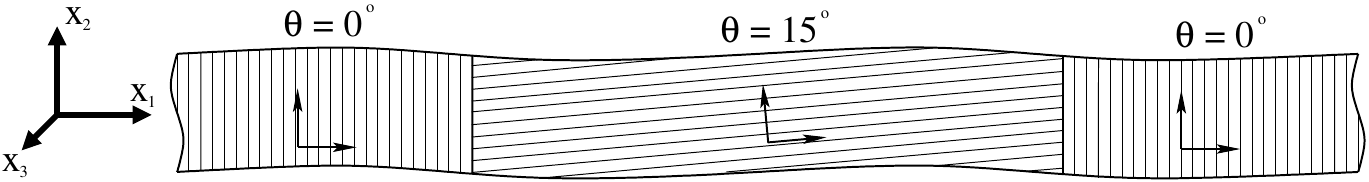}
	\caption{Representative periodic bicrystal structure with variation in $x_1$ direction.}
	\label{fig:bicrystal}
\end{figure}

Fig. \ref{fig:bicrystal} shows the two-dimensional periodic bicrystal structure used in most of the simulations. There are two grains with 15 degrees of misorientation, and the opposite surfaces along $x_1$ and $x_2$ directions are assumed periodic. Each grain has a width of 10 $\mu$m in $x_1$ direction, and a height of 2 $\mu$m in $x_2$ direction. However, due to periodic boundaries, the solution field is constant along $x_2$, so the length along $x_2$ is only for visualization purposes. Neumann boundary conditions are applied for order parameter $\eta$ and orientation $\theta$. The displacements are set to $\ubar{u}=\ubar{0}$ unless stated otherwise. In the absence of displacements, when the constraint $\pvec{e}{e}\equiv 0$ is satisfied by using a high enough Cosserat penalty parameter $\mu_c$, the coupled model should produce the same equilibrium profiles of $\eta$ and $\theta$ at grain boundaries compared to the asymptotic solution given in \ref{app:param} for pure HMP phase field. We validate this with the model parameters given in Table \ref{tab:param}.

Second order triangular elements with reduced integration are used. 
Since the solution is invariant in $x_2$ direction, only one element is used in this direction. Along $x_1$, we have tried discretization with 250, 500 and 1000 finite elements. The initialization of the solution variables is crucial to achieve initial convergence in the Newton-Raphson iterations. The orientation field $\theta$ is initialized by using a hyperbolic tangent in the form of \eqref{eqn:thetaan} with $c=20$. Using such a smooth function instead of a step function improves the convergence behavior. To make sure that the coupling function $g(\eta)$ is not singular initially, which becomes singular at $g(\eta=1)$ as seen in \eqref{eqn:fung}, the order parameter $\eta$ is initialized to a constant value of $\eta_0=0.99$. Finally, in the undeformed state, in order to make sure that there is no initial elastic strain we need to set $\accentset{\times}{e}^*(t=0)=-\theta(t=0)$ as shown in equation \eqref{eqn:eeskew}.

\begin{figure}[ht]
	\centering
	\includegraphics[width=1\textwidth]{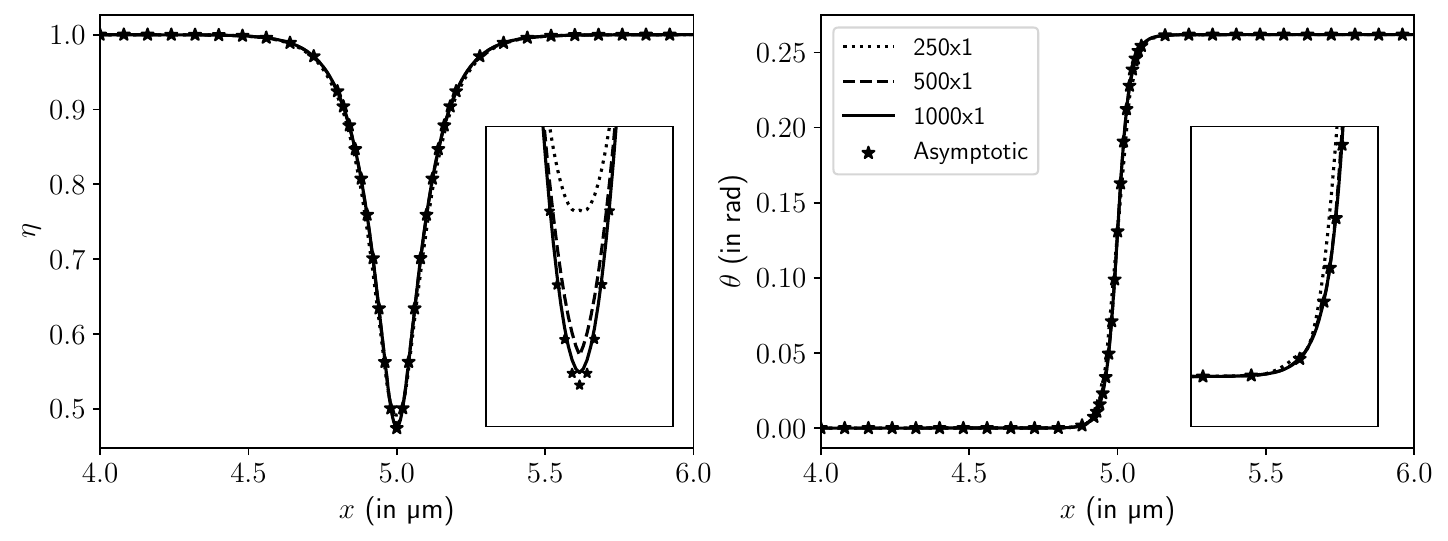}
	\caption{Equilibrium profiles of $\eta$ and $\theta$ plotted along $x_1$ direction. Results with increasing number of finite elements are shown together with analytic result. Zoomed in views on the lower-right corners reveal the small differences.}
	\label{fig:equilibrium}
\end{figure}

We let the fields evolve using time increment size $\Delta t=0.1$ s until $t=10$ seconds, at which point it reaches equilibrium. The resulting profiles are shown in Fig. \ref{fig:equilibrium} for different levels of discretization. With 1000 elements we obtained excellent agreement with the analytical solution. Even with 250 elements, the difference in minimal to the naked eye.

\begin{figure}[ht]
	\centering
	\includegraphics[width=1\textwidth]{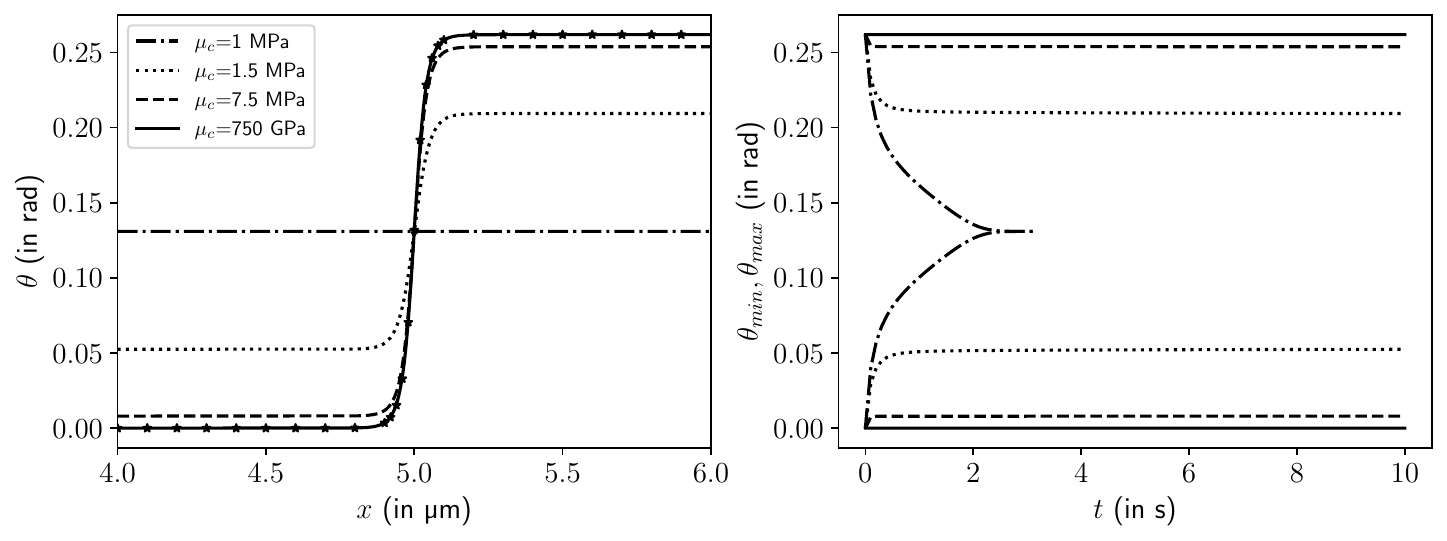}
	\caption{Equilibrium profiles $\theta$ plotted at $t=10$ seconds along $x_1$ direction for varying Cosserat penalty parameter $\mu_c$ (left), where stars represent $-\overset{\times}{e}$$^*$. Change of orientation at each grain with time (right).}
	\label{fig:thetaminmax}
\end{figure}

For the results in Fig. \ref{fig:equilibrium}, we have used $\mu_c=750$ GPa which was sufficiently high to satisfy the constraint $\pvec{e}{e}\equiv 0$. Next, we examine what happens if smaller or larger values of $\mu_c$ are used. Without displacement and slip, the only component of skew part of stress is given by

\begin{equation}
	\accentset{\times}{\sigma}=2\mu_c\accentset{\times}{e}^e=2\mu_c(-\accentset{\times}{e}^*-\theta),
\end{equation}
which is initialized to zero. When $\mu_c$ is high, to satisfy balance equation \eqref{eqn:balang}, $\accentset{\times}{e}^e$ is forced to be small. If $\mu_c$ is smaller, we would expect that $\accentset{\times}{e}^e$, thus the difference between $\accentset{\times}{e}^*$ and $\theta$ grows. The evolution of $\accentset{\times}{e}^*$ is governed by

\begin{equation}\label{eqn:estarevo}
	g(\eta)\hat{\tau}_*\accentset{\times}{\dot{e}}^*=\accentset{\times}{\sigma},
\end{equation}
which is hindered in the bulk of the grains due to the singular function $g(\eta)$. Therefore, in the case of smaller $\mu_c$, $\theta$ should move away from $\accentset{\times}{\dot{e}}^*$, meaning in our bicrystal example the two grains should rotate towards each other to minimize energy. This hypothesis is confirmed in Fig. \ref{fig:thetaminmax}, where for $\mu_c=1$ MPa the grains quickly rotate towards each other until they have the same orientation. Note that this is not the same $\theta$ relaxation mechanism as it is in the pure HMP or KWC phase field models. The difference can be seen for cases $\mu_c=1.5$ and 7.5 MPa, in which grains rotate a bit but then stop rotating when they reach equilibrium at some misorientation. This rotation decreases as $\mu_c$ increases, and it is not noticeable for $\mu_c=750$ GPa, where $\theta$ remains close to $-\accentset{\times}{e}^*$. 

\begin{figure}[ht]
	\centering
	\includegraphics[width=1\textwidth]{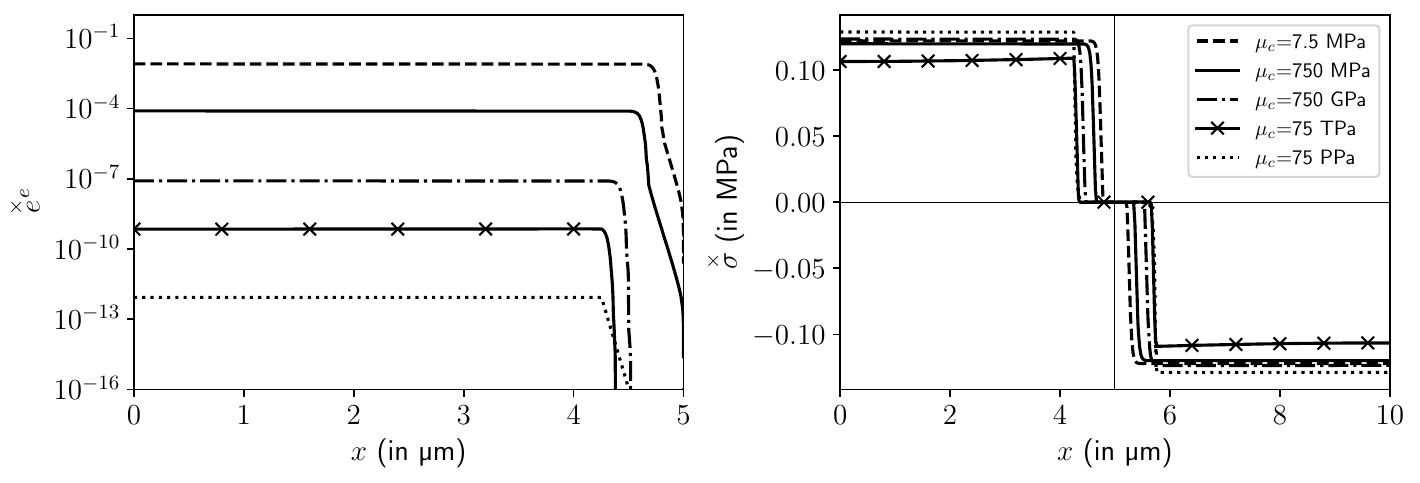}
	\caption{Skew part of elastic strain $\overset{\times}{e}$$^e$ (left) and stress $\overset{\times}{\sigma}$ plotted at $t=100$ seconds along $x_1$ direction for varying Cosserat penalty parameter $\mu_c$.}
	\label{fig:skweelsig12}
\end{figure}

\begin{figure}[ht]
	\centering
	\includegraphics[width=1\textwidth]{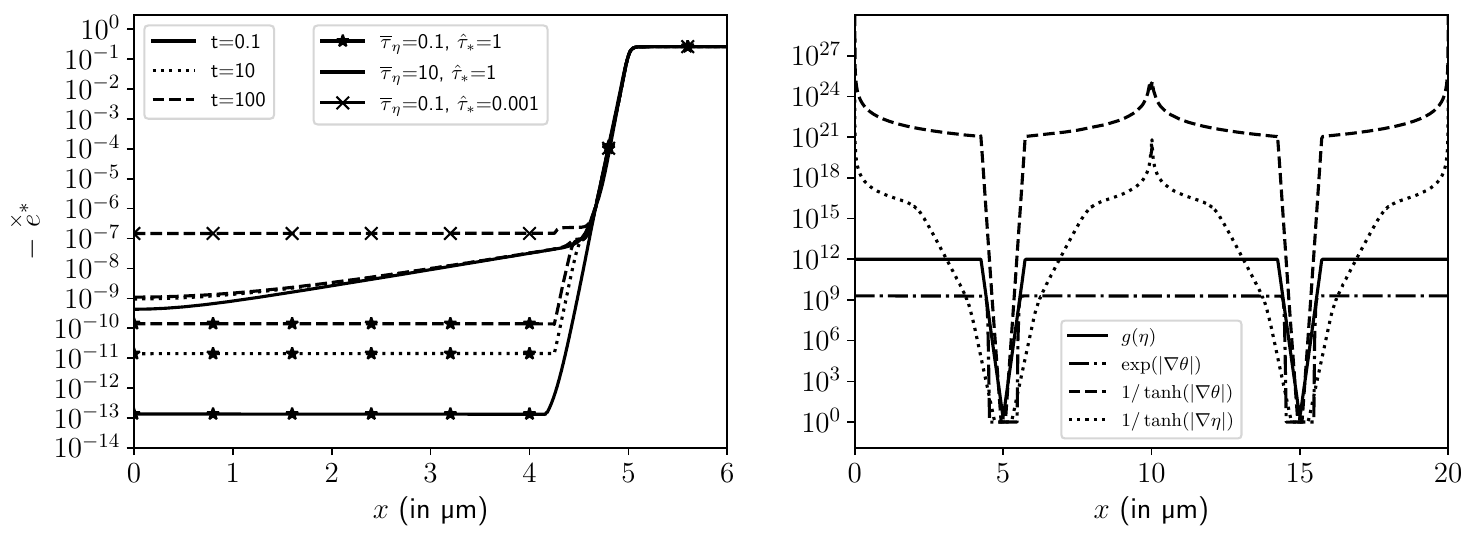}
	\caption{Eigenrotation $\overset{\times}{e}$$^*$ at different times (left) and various localization functions at $t=100$ seconds (right) plotted along $x_1$ direction. For the localizing functions \eqref{eqn:locfunc} the parameters used are $\overline{\mu}_p=10^9$, $\beta_p=10^2$, $\overline{\epsilon}=0.5$ and $\overline{C}_*=\sqrt{10}$.}
	\label{fig:fung}
\end{figure}

Fig. \ref{fig:skweelsig12} shows the difference between $-\accentset{\times}{e}^*$ and $\theta$, or, equivalently, the skew part of elastic strain $\accentset{\times}{e}^e$, for larger $\mu_c$. At $\mu_c=750$ GPa, $\accentset{\times}{e}^e$ is indeed negligible, and becomes smaller if $\mu_c$ is increased further. In fact, $\accentset{\times}{e}^e$ appears to be inversely proportional to $\mu_c$. Another observation is that the value of $\accentset{\times}{e}^e$ is many orders smaller inside the grain boundary compared to the bulk, even effectively zero for $\mu_c>750$ GPa. This is in agreement with equation \eqref{eqn:estarevo}, as $\accentset{\times}{e}^*$ evolves a lot faster inside the grain boundary and is able to remain equal to $-\theta$. Fig. \ref{fig:skweelsig12} also shows the skew part of stress $\accentset{\times}{\sigma}$ which is the driving force for $\accentset{\times}{e}^*$ evolution. Although $\mu_c$ changes significantly, the stresses $\accentset{\times}{\sigma}=2\mu_c\accentset{\times}{e}^e$ are relatively similar since $\mu_c$ and $\accentset{\times}{e}^e$ change proportionally. The non-zero stress inside the grains shows that even with very high $\mu_c$, $\accentset{\times}{e}^*$ will change in time, and as a result orientation field $\theta$ will change. However thanks to $g(\eta)$ slowing down $\accentset{\times}{e}^*$ evolution, this rotation remains negligible, $\approx 1e^{-10}$ (see Fig. \ref{fig:fung}), in given simulation time 100 seconds.

The evolution of $\accentset{\times}{e}^*$ plays a central role to equilibrium profile. It is also significant in grain boundary migration dynamics, which is discussed in Section \ref{sec:gbmigration}. Therefore, it is important understand how its evolution is localized to the grain boundaries in equation \eqref{eqn:estarevo}, and how other model parameters affect it. Fig. \ref{fig:fung} shows the change of $\accentset{\times}{e}^*$ in time with $\mu_c=750$ GPa for two different inverse mobility parameters $\overline{\tau}_\eta$: 0.1 and 10. At $t=0$, $\accentset{\times}{e}^*$ is initialized to zero in the outer grain, and $t=0.1$ is the first increment, where the value of $\accentset{\times}{e}^*$ is different by several orders of magnitude for the two cases. As time passes, the eigenrotation keeps increasing in both cases, which is caused by the skew-symmetric stress in Fig. \ref{fig:skweelsig12}. Also, for $\overline{\tau}_\eta=10$, unlike $\overline{\tau}_\eta=0.1$ the eigenrotation is not constant inside the bulk. This difference is due to our localizing singular function $g(\eta)$ in \eqref{eqn:estarevo}. When $\overline{\tau}_\eta$ is increased, $\eta$ evolves slower, which is initialized as 0.99, meaning $g(\eta)$ is not as singular as $g(1)$ initially. Thus, when $\eta$ evolves slowly toward 1, $g(\eta)$ fails to prevent $\accentset{\times}{e}^*$ evolution inside the grains. Therefore, $\overline{\tau}_\eta$ should be chosen sufficiently small such as $\overline{\tau}_\eta=0.1$. Another option is to initialize $\eta$ closer to 1, i.e. 0.9999, which however makes the initial convergence harder. Looking at equation \eqref{eqn:estarevo}, the effect of $\hat{\tau}_*$ on $\accentset{\times}{\dot{e}}^*$ is clear, as $\hat{\tau}_*$ decreases $\accentset{\times}{e}^*$ evolves more rapidly in a linearly proportional way. This is observed in Fig. \ref{fig:fung} when $\hat{\tau}_*=0.001$ is used instead of $\hat{\tau}_*=1$. Consequently, using a larger $\hat{\tau}_*$ seems beneficial for equilibrium purposes, but this has a negative effect on grain boundary mobility as discussed in Section \ref{sec:gbmigration}.

Another important factor in $\accentset{\times}{e}^*$ evolution is the function that localizes the evolution to grain boundaries. As mentioned in previous section, the KWC-CCP coupled model is presented with different localizing functions shown in \eqref{eqn:locfunc} throughout the development (\cite{ask2018cosserat,ask2019cosserat,ask2020microstructure}). In our proposed model, we use singular the coupling function $g(\eta)$ instead, which is the cornerstone of the HMP phase field model (\cite{henry2012orientation,staublin2022phase}). These different localizing functions are plotted and compared in Fig. \ref{fig:fung}. Notice that the $g(\eta)$ shown actually has a finite value of $10^{12}$ inside the grains. This is because we are using a predefined cut-off value of ($1-10^{-4}$) to calculate $g$ and $g'$ during numerical simulations (see \ref{app:fenics}), which improves stability and convergence behavior of the model significantly. Similarly, \cite{ask2020microstructure} have used a smoother and less strict exponential form compared to previous $\tanh$ forms in \cite{ask2018cosserat,ask2019cosserat}. The $\tanh(|\nabla\eta|)$ form has a wider distribution around grain boundary which is not ideal for localization. The $\tanh(|\nabla\theta|)$ form has a similar distribution compared to $g(\eta)$, but it takes very large values inside the grains. In our simulations we have observed that $\tanh$ forms would result in peaks of $\accentset{\times}{e}^e$ near grain boundaries, while the function $g(\eta)$ provides a good balance between smoothness and localization; moreover we can control its value inside the grains with the cut-off value.

\subsection{Triple-junction test}

\begin{figure}[ht]
	\centering
	\includegraphics[width=0.8\textwidth]{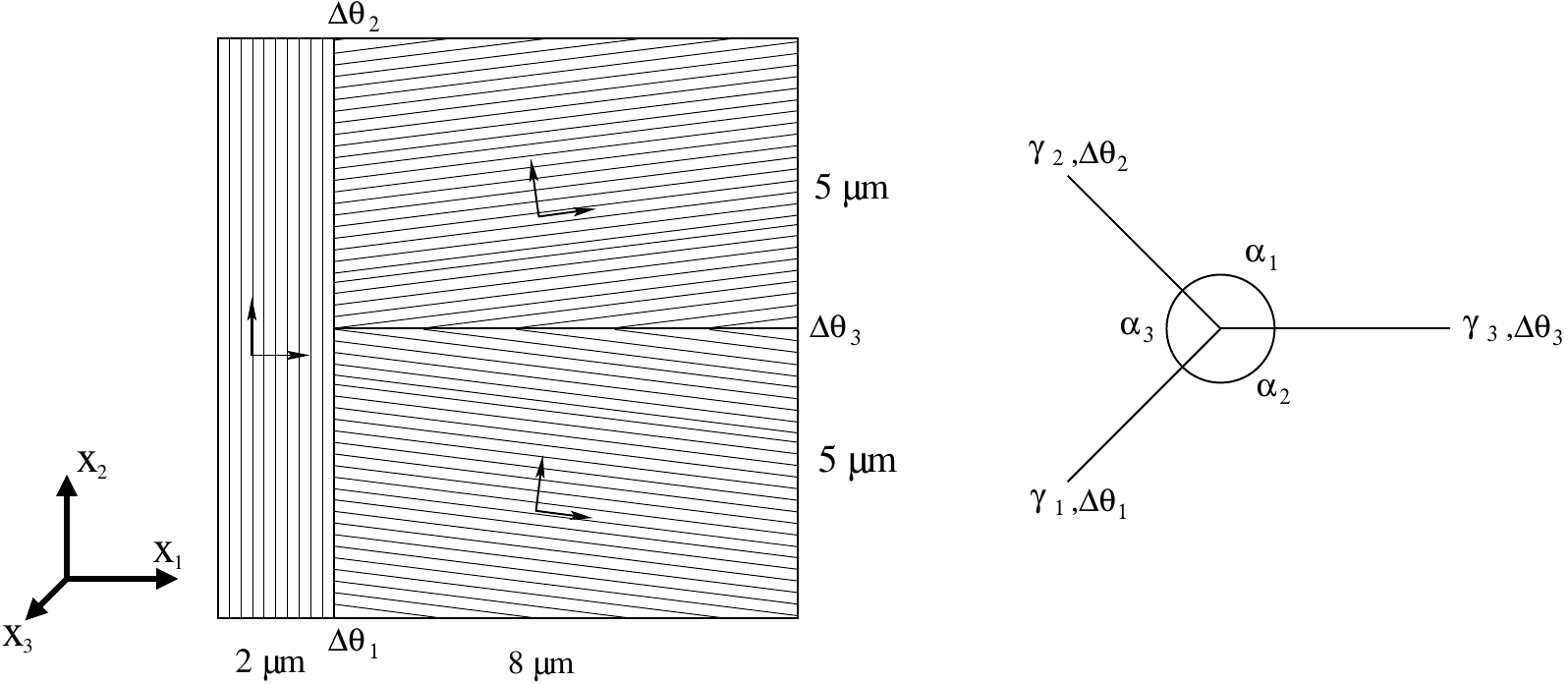}
	\caption{Triple junction test geometry and the dihedral angles at the junction.}
	\label{fig:tripjunc}
\end{figure}

In this section we perform the triple-junction test with the geometry shown in Fig. \ref{fig:tripjunc}. The grains have different orientations and the corresponding misorientations at the grain boundaries are given by $\Delta\theta_1$, $\Delta\theta_2$ and $\Delta\theta_3$. The orientations are held constant by applying Dirichlet boundary conditions on $\eta$ and $\theta$. Under these conditions, the triple junction moves to an equilibrium point and stops. The angles of the intersection at the equilibrium are given by Herring's force balance \citep{herring1951some}, which simplifies to Young's Law for isotropic grain boundaries

\begin{equation}\label{eqn:herring}
	\dfrac{\gamma_1}{\sin(\alpha_1)}=\dfrac{\gamma_2}{\sin(\alpha_2)}=\dfrac{\gamma_3}{\sin(\alpha_3)},
\end{equation}
where $\gamma_i$ are grain boundary energies and $\alpha_i$ are dihedral angles. We test the proposed coupled model by using different misorientations with parameters in Table \ref{tab:param}, to see if correct dihedral angles from \eqref{eqn:herring} are obtained. While this test has been already performed in previous work (\cite{henry2012orientation,staublin2022phase}) for the original HMP orientation phase field, we use a different potential $V(\eta)$ and evolution is governed by Cosserat mechanics.

The geometry is divided into 100x100 blocks where each block contains four second order triangular finite elements with reduced integration. The simulations are continued until time-independent solutions are obtained. The triple junction angles from Young's Law \eqref{eqn:herring} and the angles from finite element results are compared in Table \ref{tab:tripjunc}. The grain boundary energies used in \eqref{eqn:herring} are found from 1D grain boundary simulations for the given discretization. The angles are measured along the lines of minimum $\eta$, i.e., the center of grain boundary.

\begin{table}[ht]
	\caption{Equilibrium triple-junction angles obrained from finite element simulations and Herring's equation.}\label{tab:tripjunc}
	\centering
	\begin{tabular}{ccc|ccc|ccc}
		\multicolumn{3}{c|}{Misorientation}                    & \multicolumn{3}{c|}{GB energy (in J/m$^{-2}$)} & \multicolumn{3}{c}{FE angles (Herring)}                               \\ \hline
		$\Delta\theta_1$ & $\Delta\theta_2$ & $\Delta\theta_2$ & $\gamma_1$     & $\gamma_2$     & $\gamma_3$    & $\alpha_1$            & $\alpha_2$            & $\alpha_3$            \\ \hline
		5$^o$            & 5$^o$            & 10$^o$           & 0.2435         & 0.2435         & 0.3914        & 143.6$^o$ (143.5$^o$) & 143.6$^o$ (143.5$^o$) & 72.8$^o$ (73.0$^o$)   \\
		7.5$^o$          & 7.5$^o$          & 15$^o$           & 0.3235         & 0.3235         & 0.5018        & 141.0$^o$ (140.9$^o$) & 141.0$^o$ (140.9$^o$) & 78.0$^o$ (78.2$^o$)   \\
		10$^o$           & 10$^o$           & 20$^o$           & 0.3914         & 0.3914         & 0.5883        & 138.8$^o$ (138.7$^o$) & 138.8$^o$ (138.7$^o$) & 82.4$^o$ (82.6$^o$)   \\
		20$^o$           & 20$^o$           & 40$^o$           & 0.5883         & 0.5883         & 0.8038        & 132.2$^o$ (133.1$^o$) & 132.2$^o$ (133.1$^o$) & 95.6$^o$ (93.8$^o$)   \\
		30$^o$           & 30$^o$           & 60$^o$           & 0.7154         & 0.7154         & 0.9146        & 130.1$^o$ (129.7$^o$) & 130.1$^o$(129.7$^o$)  & 99.8$^o$ (100.6$^o$)  \\
		20$^o$           & 15$^o$           & 5$^o$            & 0.5883         & 0.5018         & 0.2435        & 80.4$^o$ (81.8$^o$)   & 122.7$^o$ (122.4$^o$) & 156.9$^o$ (155.8$^o$) \\
		10$^o$           & 5$^o$            & 15$^o$           & 0.3914         & 0.2435         & 0.5018        & 130.6$^o$ (130.3$^o$) & 151.6$^o$ (151.7$^o$) & 77.8$^o$ (78.0$^o$)  
	\end{tabular}
\end{table}

\begin{figure}[ht!]
	\centering
	\includegraphics[width=0.8\textwidth]{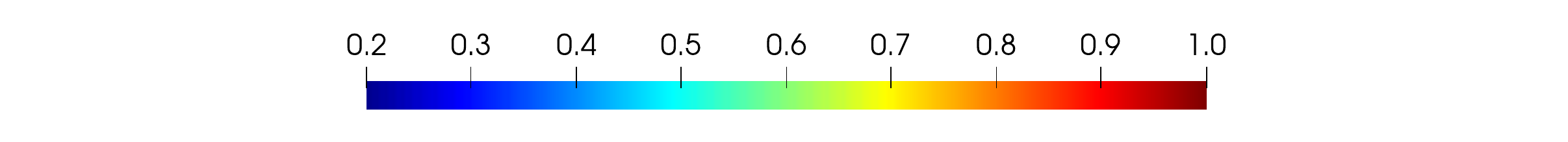}
	\includegraphics[width=0.24\textwidth]{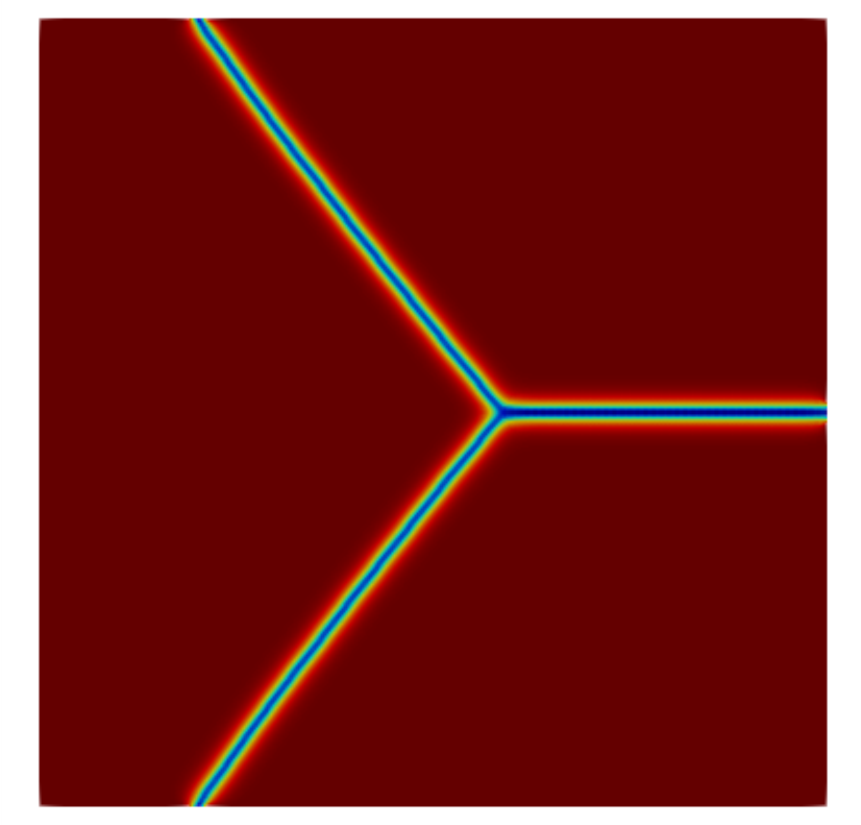}
	\includegraphics[width=0.24\textwidth]{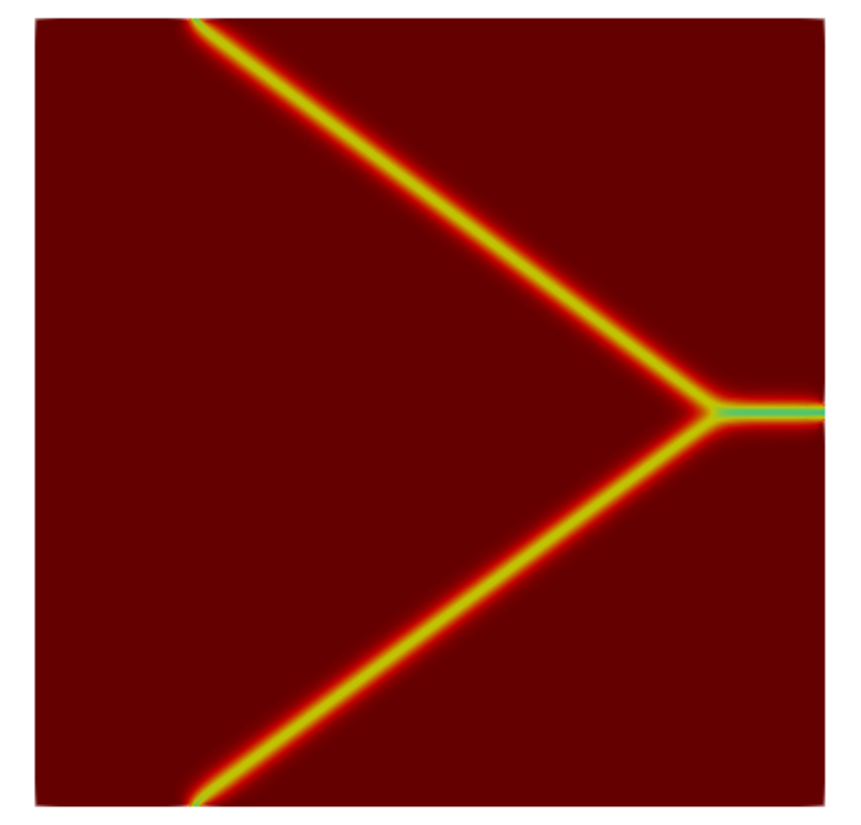}
	\includegraphics[width=0.24\textwidth]{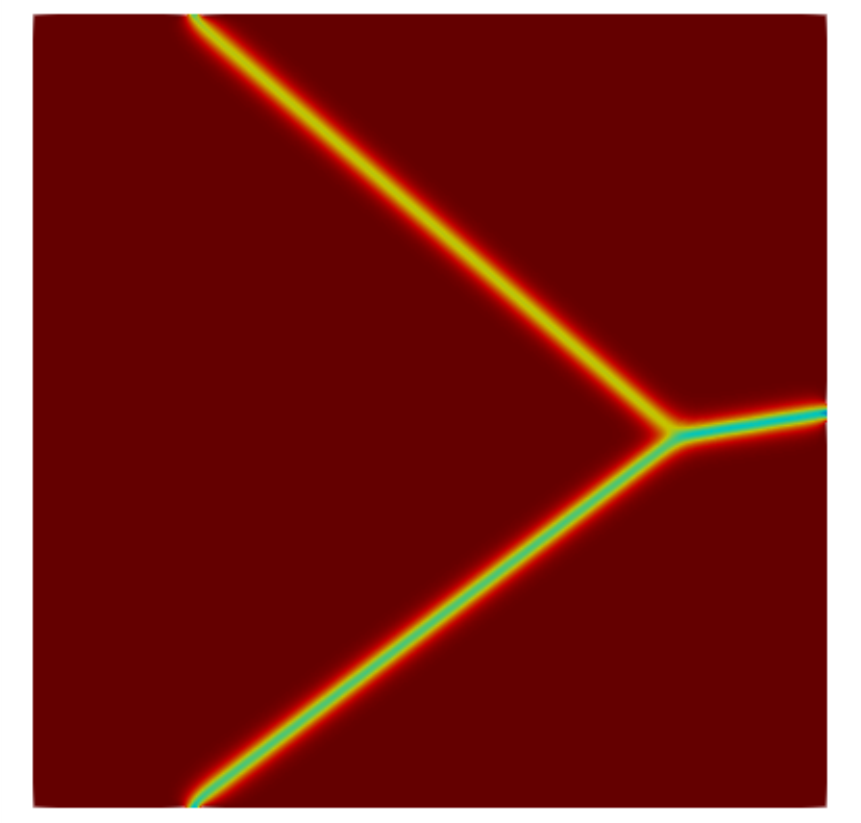}
	\includegraphics[width=0.24\textwidth]{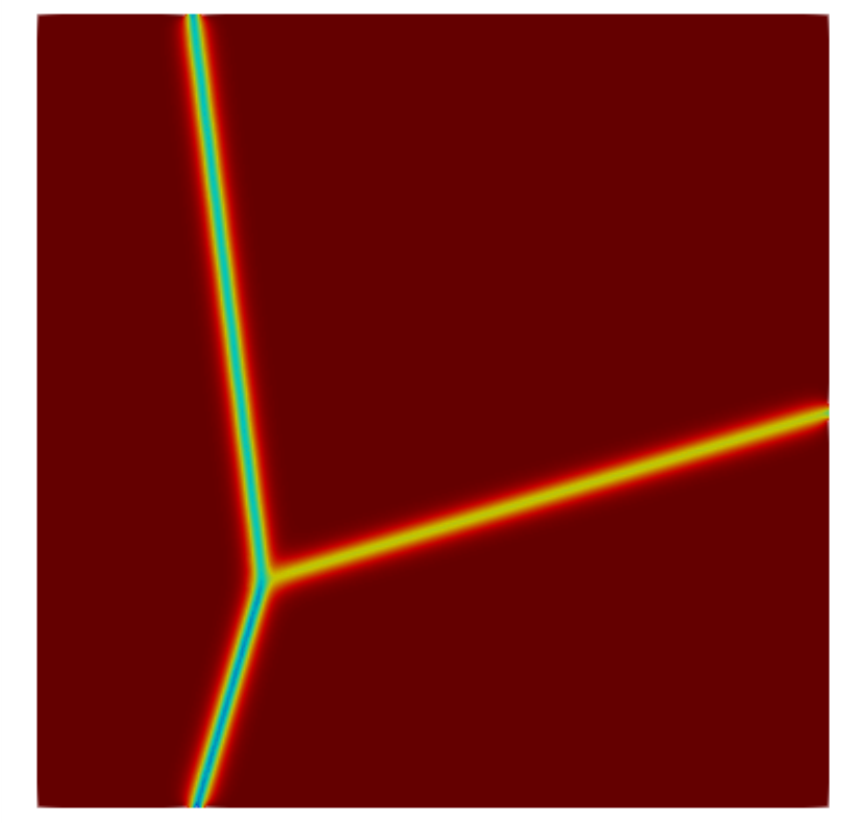}
	\caption{Order parameter contours at equilibrium of triple junction problem for different misorientations, $\Delta\theta_1,\Delta\theta_2,\Delta\theta_3$, from left to right: 30$^o$/30$^o$/60$^o$, 5$^o$/5$^o$/10$^o$, 10$^o$/5$^o$/15$^o$ and 20$^o$/15$^o$/5$^o$.}
	\label{fig:tripjunceta}
\end{figure}

As seen in Table \ref{tab:tripjunc}, the agreement is very good as expected of phase field models. Fig. \ref{fig:tripjunceta} shows the $\eta$ fields after equilibrium is reached where the equilibrium positions are clearly different. With these we confirm that mechanics coupled phase field model is working as intended.

\subsection{Elastic shear loading}

In this section we test the behavior of the model under mechanical loading. The same periodic bicrystal structure as in Section \ref{sec:equi} is used with 1000 blocks of second order triangular elements with reduced integration along $x_1$ direction. Neumann boundary conditions are employed for $\eta$ and $\theta$. The loading is applied by imposing a mean displacement field with periodic fluctuations as explained in \ref{app:shearan}. The total displacement $\underline{\bm{u}}$ is given by,

\begin{equation}
	\underline{\bm{u}}=\tend{B}\cdot\underline{\bm{x}}+\underline{\bm{v}}
\end{equation}
where the tensor $\tend{B}$ is defined as,

\begin{equation}
	\tend{B}=
	\begin{bmatrix}
		0 & B_{12} & 0\;\;\\
		B_{21} & 0 & 0\\
		0 & 0 & 0
	\end{bmatrix}
\end{equation}
for shear loading with $B_{21}=0.001$ and $B_{12}=0$. The periodic fluctuation $\underline{\bm{v}}$ is fixed at the corners. The loading is applied until $t=10$ s with $\Delta t=0.1$ s. In order to compare our numerical results we assume that the material is elastic since it is possible to find analytic solution for this problem. We consider two test cases, isotropic and cubic elasticity.

\subsubsection{Isotropic elasticity}
\begin{figure}[ht]
	\centering
	\includegraphics[width=1\textwidth]{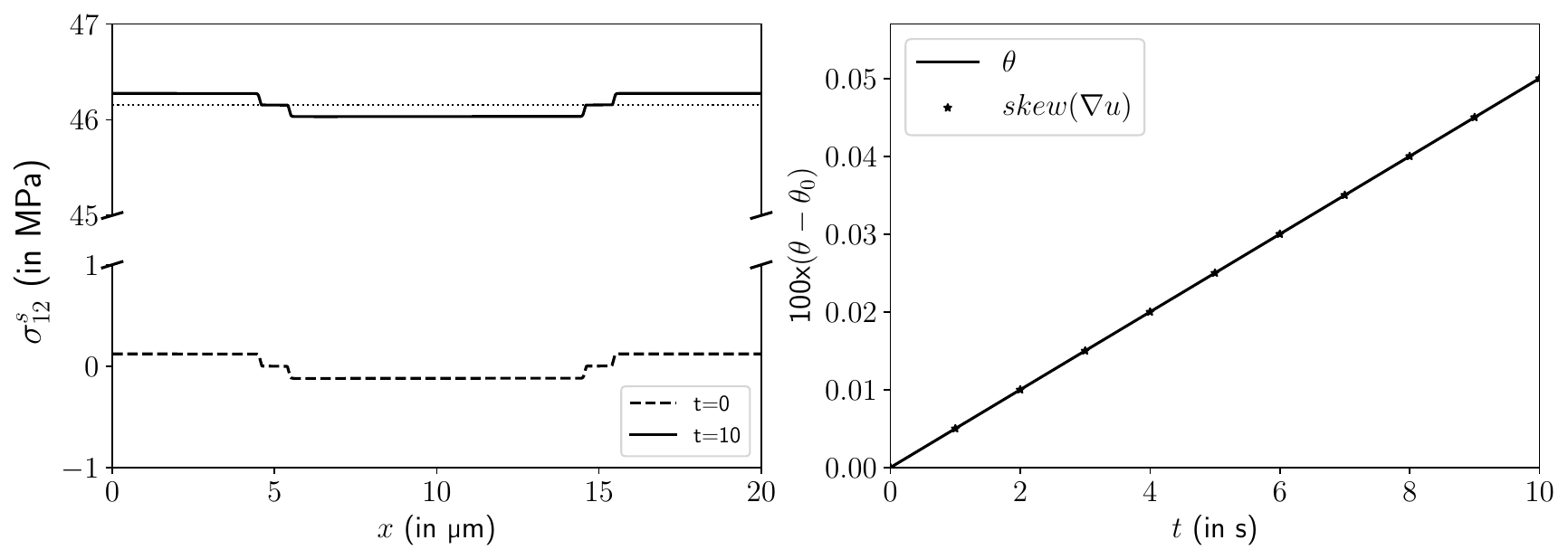}
	\caption{Stress profile before and after simple shear loading for isotropic elasticity, where dotted line shows analytic solution (left). Cosserat rotation and lattice rotation scaled by 100 (right).}
	\label{fig:sheariso}
\end{figure}

For the isotropic case the model parameters given in Table \ref{tab:param} are used except for the elastic constants, for which we assume Young's modulus $E=120$ GPa and Poisson's ratio as $\nu=0.3$. Before applying the loading, we first initialize the order parameter and orientation field similar to Section \ref{sec:equi} and solve until reaching an equilibrium. However, unlike the first example, the displacement degree of freedom is not restricted, i.e. $\ubar{u}\not\equiv 0$. The obtained fields $\eta$, $\theta$, $\pvec{e}{*}$ and $\ubar{u}$ are used as initial conditions in the loading stage. When $\ubar{u}$ is not fixed, we have an initial symmetric stress distribution at $t=0$ as seen in Fig. \ref{fig:sheariso}, where $\tend{\sigma}^s=\tenq{E}^s:\tend{\varepsilon}^e$, in addition to skew-symmetric stress $\pvec{\sigma}{}$. This initial symmetric stress will be referred to as a residual stress from here on. The reason for the residual stress is the tendency of the grains to rotate towards each other. In the HMP or KWC orientation phase field models, one of the evolution mechanisms to minimize free energy is to reduce misorientation by the rotation of grains. This rotation can be suppressed by using a varying mobility function such as \eqref{eqn:locfunc} or \eqref{eqn:taustarg}. In the coupled model, the magnitude of the driving force for this rotation is given by $\pvec{\sigma}{}$ as shown in Fig. \ref{fig:skweelsig12}. Hence, when displacements are free in Fig. \ref{fig:bicrystal}, outer grain rotates counter-clockwise and inner grain rotates clockwise. Thus, in order to satisfy continuity at the grain boundaries, both grains must be strained which we observe as the residual stress in Fig. \ref{fig:sheariso}. The residual stress is zero in the vicinity of grain boundaries since it is relaxed quickly according to \eqref{eqn:wstar} by the evolution of $\pvec{e}{*}$. Since the magnitude of residual stress is related to the driving force for grain rotation, it is smaller for higher misorientations, and for grain boundaries that are further away. In our example it is quite small, approximately 0.125 MPa, and it is magnified in Fig. \ref{fig:sheariso} for clarity.

For the isotropic case, the stress distribution after loading should be the same in both grains since there is no orientation dependence. The analytic solution is straightforward and given by $\sigma_{12}^s=46.15$ MPa with other components zero, which is shown as dotted line in Fig. \ref{fig:sheariso}. The final stress from the simulation is the same as analytic solution if residual stress is excluded. On the right of Fig. \ref{fig:sheariso}, the change of lattice rotation $skew(\nabla u)$ and Cosserat rotation $\theta$ are shown as functions of time. They are equal as desired and change linearly as expected. The coupled model allows the continuous orientation field $\theta$ to evolve due to mechanical loading.

\subsubsection{Cubic elasticity}
\begin{figure}[ht]
	\centering
	\includegraphics[width=1\textwidth]{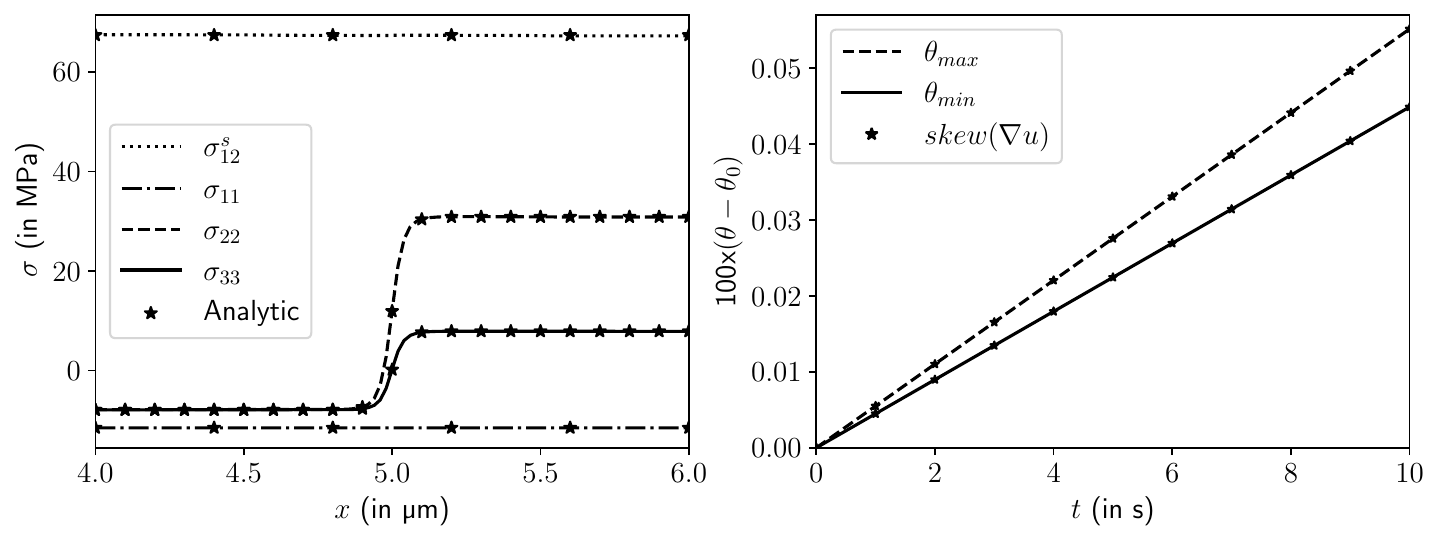}
	\caption{Stress profiles after simple shear loading at $t=10$ for cubic anisotropy, where stars show analytic solution (left). Cosserat rotation and lattice rotation scaled by 100 for grains with $0^o$ ($\theta_{min}$) and $15^o$ ($\theta_{max}$) orientation (right).}
	\label{fig:shearcub}
\end{figure}

In the case of cubic anisotropy, elastic constants are taken as $C_{11}=160$ GPa, $C_{12}=110$ GPa and $C_{44}=75$ GPa. Due to anisotropy and different orientations of grains, the final stress field includes components other than shear as shown in Fig. \ref{fig:shearcub}. For the given boundary conditions, it is possible to find an analytic solution as explained in \ref{app:shearan}. The analytic results are plotted with markers on top of the numerical solution in Fig. \ref{fig:shearcub}, showing excellent agreement. Note that, similar to isotropic case, we have small residual stresses after initializing the order parameter and orientation field. Fig. \ref{fig:shearcub} (right) shows again the lattice and Cosserat rotations together, but this time they are different at each grain due to anisotropy. As expected, for high Cosserat penalty parameter $\mu_c$, the Cosserat rotation $\theta$ represents the lattice rotation $skew(\nabla u)$.

\subsection{Grain boundary migration due to stored dislocations}\label{sec:gbmigration}

In this section the mechanism of grain boundary migration driven by statistically stored dislocations (SSD) is discussed. During plastic deformation of a polycrystal, SSDs are most likely generated non-uniformly, which causes a stored energy gradient. In the free energy \eqref{eqn:freeen2d}, assuming single slip system for simplicity this is given by the term

\begin{equation}
	\psi_\rho(\eta,r)=\phi(\eta)\frac{\lambda}{2}\mu^eb^2\rho,
\end{equation}
which can cause an energy gradient depending on the distribution of $\rho$. Moreover, in the evolution of the phase field, due to the term $\phi(\eta)$, this results in the driving force

\begin{equation}
	\psi_{\rho,\eta}(\eta,r)=\phi_{,\eta}(\eta)\frac{\lambda}{2}\mu^eb^2\rho.\label{eqn:ssden}
\end{equation}
We test this mechanism using the same periodic bicrystal example in Fig. \ref{fig:bicrystal}. The SSD density $\rho_0$ is initialized as $10^{15}$ m$^{-2}$ in the inner grain resembling cold worked copper \citep{rollett2017recrystallization} and as zero in the outer grain. The shear modulus $\mu^e$ is given by $C_{44}=75$ GPa for cubic elasticity, and the Burgers vector for pure copper is $b=0.2556$ nm and $\lambda=0.3$ (see Eq. \ref{eqn:ssden}). The mobility constant for $\pvec{e}{*}$ is taken as $\overline{\tau}_*=0.01$ and, to counter the increased mobility, the cut-off of $g(\eta)$ taken as $(1-10^{-5})$. The rest of the parameters are given in Table \ref{tab:param}. The same mesh as in previous example is used with Neumann boundary conditions for $\eta$ and $\theta$, but displacements $\ubar{u}$ are set to zero. The solution fields are initialized with the equilibrium values from Section \ref{sec:equi}.

\begin{figure}[ht]
	\centering
	\includegraphics[width=1\textwidth]{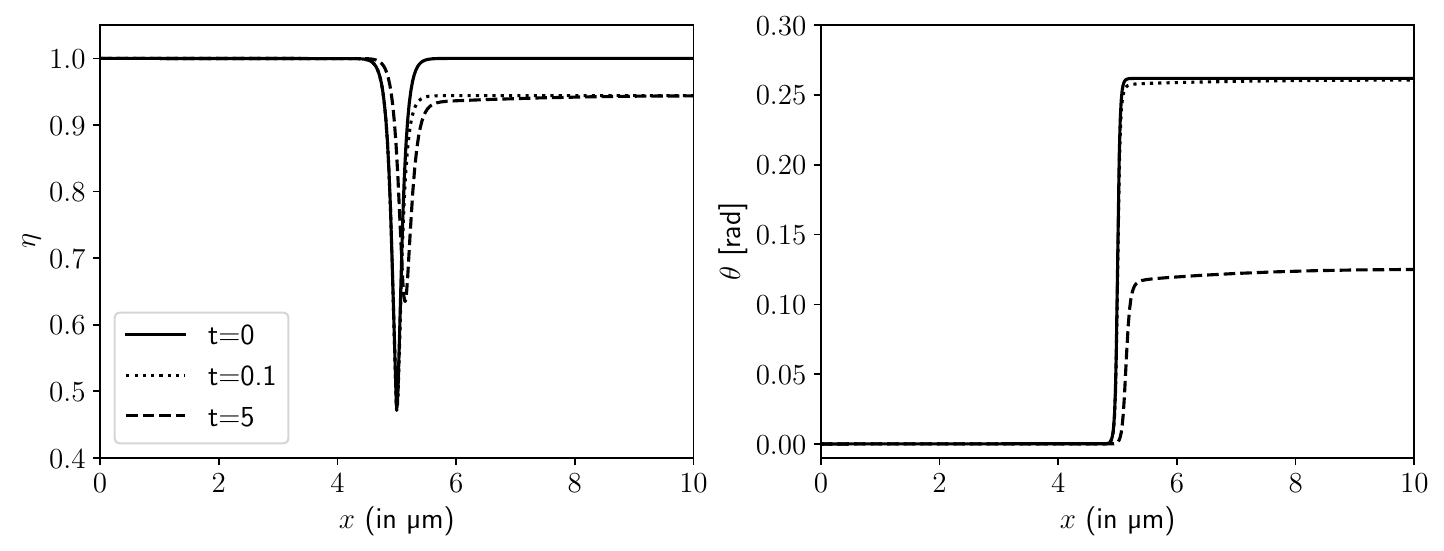}
	\caption{Profiles of $\eta$ and $\theta$ at different times when SSD energy multiplier function is $\phi(\eta)=\eta$, and inner grain has stored SSD density of $\rho=10^{15}$ $m^{-2}$.}
	\label{fig:gbmigphi0}
\end{figure}

\begin{figure}[ht]
	\centering
	\includegraphics[width=1\textwidth]{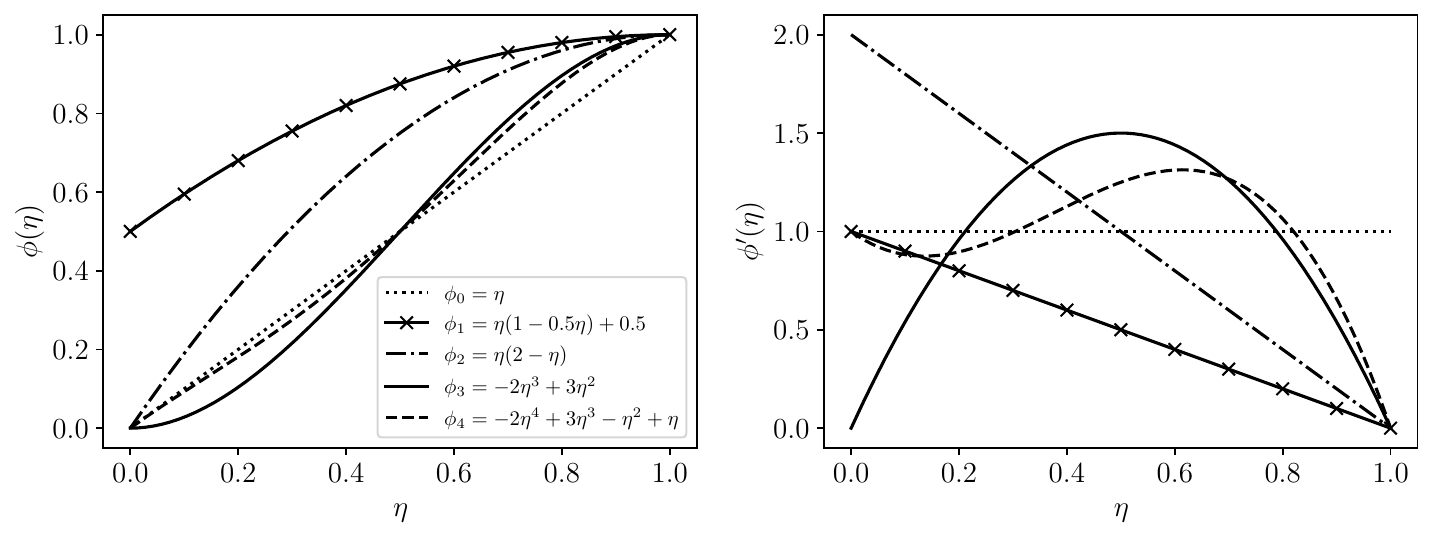}
	\caption{Different possible forms of SSD energy multiplier function $\phi(\eta)$ and their derivatives $\phi'(\eta)$.}
	\label{fig:gbmigphi}
\end{figure}

We first used $\phi(\eta)=\eta$ following \cite{abrivard2012aphase} and \cite{ask2018cosserat}, which does not work for the HMP phase field coupled model as seen in Fig. \ref{fig:gbmigphi0}. While the grain boundary moves towards the grain with stored SSDs, the initial orientation of the inner grain is not preserved. This is a consequence of the way \cite{abrivard2012aphase} formulated $\psi_{\rho,\eta}(\eta,r)$; the order parameter $\eta$ stabilizes at a value $\eta<1$ given by

\begin{equation}\label{eqn:etaeq2}
	\eta^{eq}=1-\dfrac{\phi_{,\eta}(\eta^{eq})\psi_\rho(r)}{f_0\alpha},
\end{equation}
which results in $\eta^{eq}\approx 0.944$ for the given parameters, as seen in Fig. \ref{fig:gbmigphi0}. This is incompatible with the HMP model, since it relies on the singular function $g(\eta)$ to localize and distinguish grain boundaries and bulk of grains, and $\eta$ must be close to 1 for $g(\eta)$ to be singular. 
Therefore, we propose a modified $\phi(\eta)$ which lets $\eta^{eq}$ to stay at 1 even if SSDs are present. Looking at \eqref{eqn:etaeq2}, this is possible if

\begin{equation}
	\phi_{,\eta}(1)=0 \quad\text{and}\quad \phi(\eta)>0 \quad\text{for}\quad 0<\eta<1
\end{equation}
where second condition is for the positivity of energy although it is not mandatory. Several polynomial forms satisfying these conditions are shown in Fig. \ref{fig:gbmigphi} along with their derivatives, representing the driving force on the order parameter.

\begin{figure}[ht!]
	\centering
	\includegraphics[width=1\textwidth]{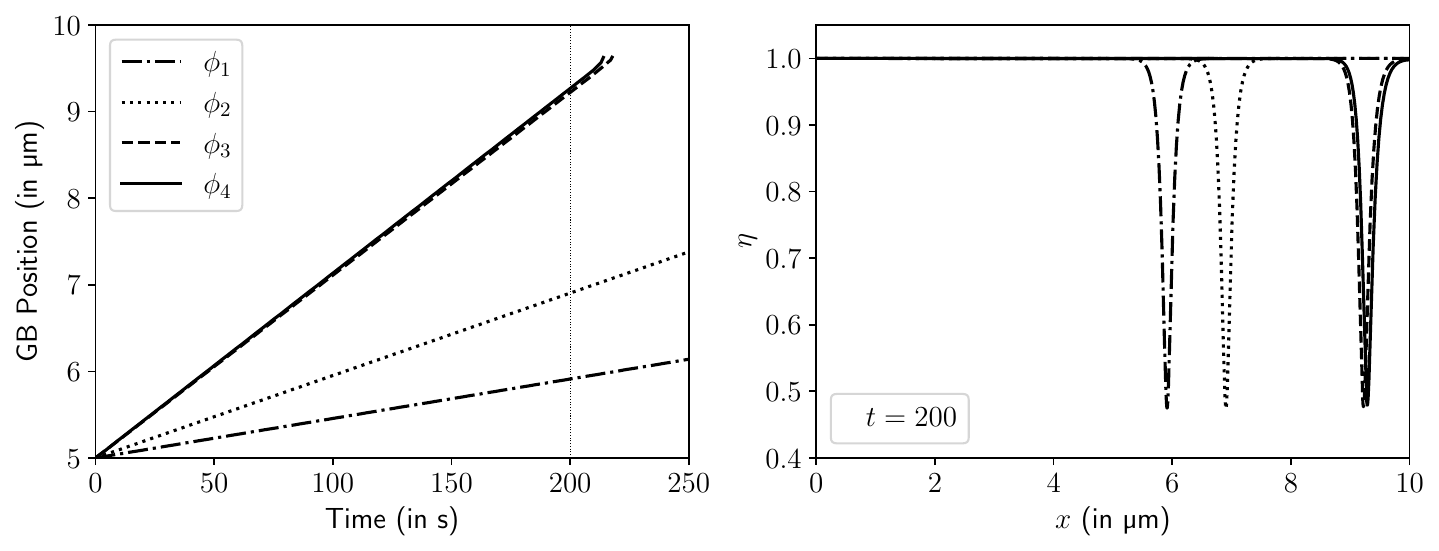}
	\caption{Grain boundary position in time for different forms of SSD energy multiplier function $\phi(\eta)$ (left), and order parameter $\eta$ profile for each at $t=200$ seconds (right).}
	\label{fig:gbmigphifunc}
\end{figure}

Fig. \ref{fig:gbmigphifunc} shows the motion of the grain boundary for each of the polynomials in Fig. \ref{fig:gbmigphi}. Left figure shows the position of the grain boundary in time; clearly, the velocity differs depending on the form of $\phi(\eta)$ with $\phi_3$ and $\phi_4$ being very similar. The right figure shows the $\eta$ profiles at $t=200$ s, where we can see that the $\eta^{eq}=1$ condition is satisfied for all $\phi(\eta)$. The forms of $\phi_3$ and $\phi_4$ are chosen to be as similar as possible to $\phi_0=\eta$. In our tests comparing the different forms of $\phi$ using the KWC-Cosserat coupled model, we have observed that $\phi_3$ and $\phi_4$ provide comparable mobility to $\phi_0=\eta$, while $\phi_1$ and $\phi_2$ significantly reduce mobility. This mobility appears to be proportional to the value of $\phi'(\eta)$ at $\eta>0.5$ shown in Fig. \ref{fig:gbmigphi} (right). This is consistent with the fact that the grain boundary is represented with order parameter $0.5<\eta<1$ for given misorientation as seen in Fig. \ref{fig:equilibrium}. Therefore, in remaining examples we use the 4th order polynomial form $\phi_4(\eta)$ which is most similar to $\phi_0=\eta$.

\begin{figure}[ht!]
	\centering
	\includegraphics[width=1\textwidth]{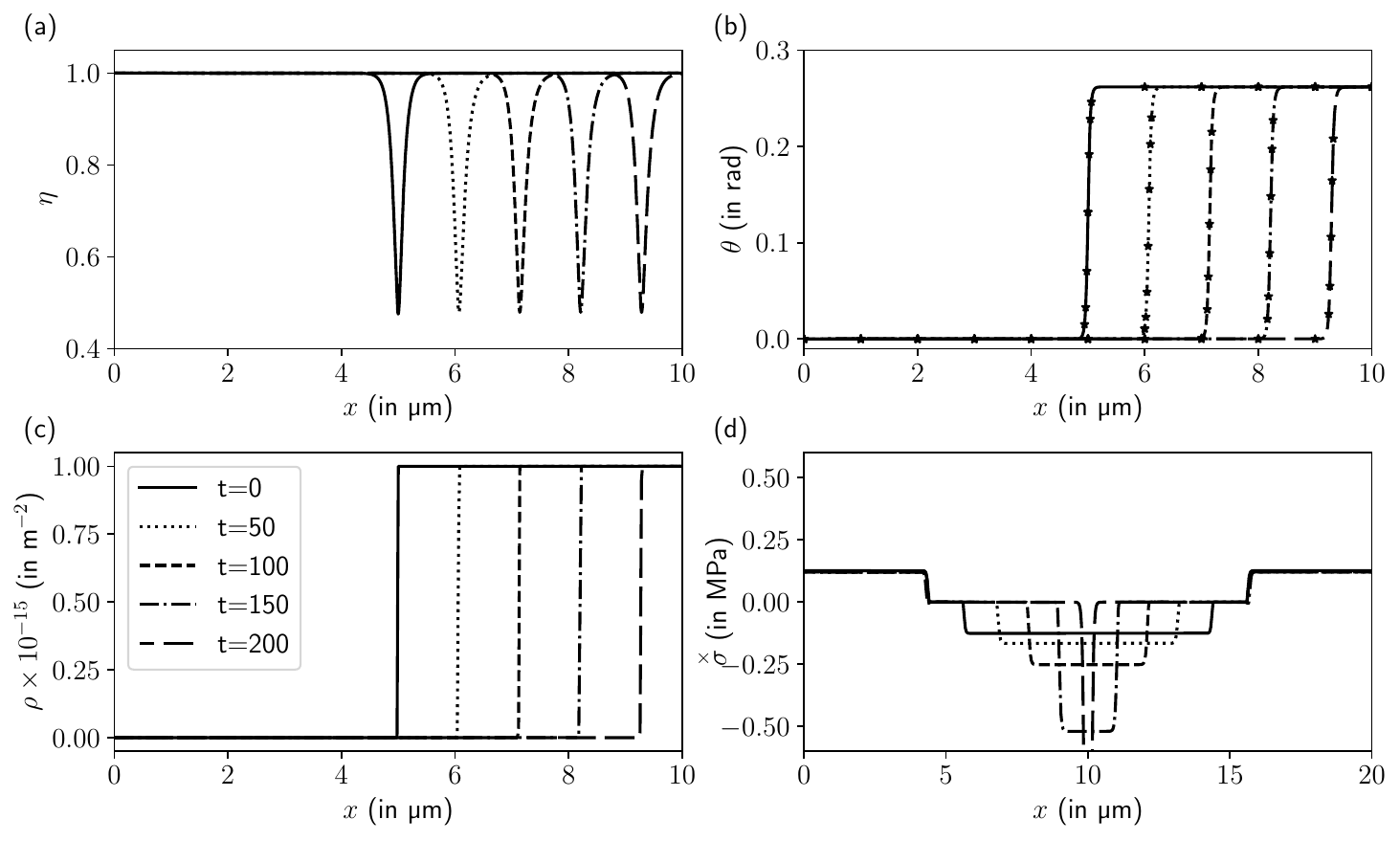}
	\caption{Grain boundary migration at different times, where SSD energy multiplier $\phi_4$ is used with $\overline{\tau}_*=0.01$. Profiles of phase field $\eta$ (a), orientation $\theta$ and $-\overset{\times}{e}$$^*$ as stars (b), SSD density $\rho$ (c) and skew-stress $\overset{\times}{\sigma}$ (d) are shown.}
	\label{fig:gbmig}
\end{figure}

Fig. \ref{fig:gbmig} shows the profiles of $\eta$, $\theta$, $\pvec{e}{*}$, $\rho$ and $\accentset{\times}{\sigma}$ at different times. At $t=0$, the inner grain is initialized with $\rho_0=10^{15}$, as seen in Fig. \ref{fig:gbmig} (c), while $\eta$, $\theta$ and $\accentset{\times}{\sigma}$ have the respective equilibrium profiles. $\accentset{\times}{\sigma}$ is non-zero inside the grains, as discussed in Section \ref{sec:equi}, showing the tendency to rotate, but held in place by a high Cosserat penalty parameter $\mu_c$. As the simulation is run, the energy gradient due to the inhomogeneous SSD distribution drives the grain boundaries to move towards the grain with the higher energy. In the wake of the moving grain boundary, static recovery can take place, which is a phenomenon observed in experiments \citep{bailey1962recrystallization}. In the model this is implemented with equation \eqref{eqn:rhodot} following \cite{abrivard2012aphase}. If the recovery parameter $C_D$ in \eqref{eqn:rhodot} is sufficiently large, i.e. 100, full recovery occurs and $\rho$ reduces to zero in the wake of the sweeping boundary, as seen in Fig. \ref{fig:gbmig} (c). 

\begin{figure}[ht!]
	\centering
	\includegraphics[width=0.45\textwidth]{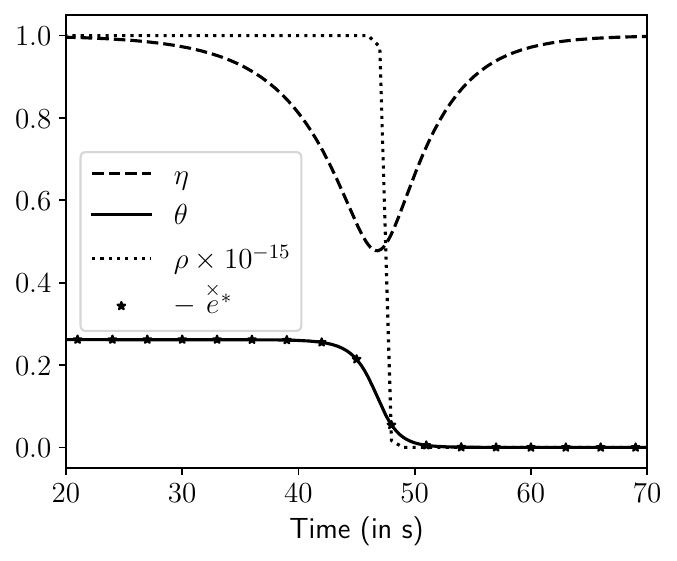}
	\caption{Evolution of solution variables in time on a point located at $x=6$ $\mu$m in Fig. \ref{fig:gbmig}}
	\label{fig:gbmigxloc}
\end{figure}

In Fig. \ref{fig:gbmig} (b) we see that the orientation field $\theta$ evolves with the moving grain boundary, and the reference orientation $\pvec{e}{*}$ follows the movement according to equation \eqref{eqn:estarevo}, ensuring that the strain free state is maintained. Moreover, the evolution of $\pvec{e}{*}$ inside the sweeping grain boundary fully relaxes the initial skew stress $\accentset{\times}{\sigma}$ in the wake as seen in Fig. \ref{fig:gbmig} (d). Also, we see an increase of $\accentset{\times}{\sigma}$ in the inner grain as the grain boundaries move closer, which shows that there is a stronger tendency to rotate as grain size decreases. Fig. \ref{fig:gbmigxloc} shows the evolution of main fields in time at the fixed location $x_1=6$ $\mu$m. The SSD density smoothly reduces to zero as discussed by \cite{abrivard2012aphase}, while $\theta$ and $\pvec{e}{*}$ evolve together.

\begin{figure}[ht!]
	\centering
	\includegraphics[width=1\textwidth]{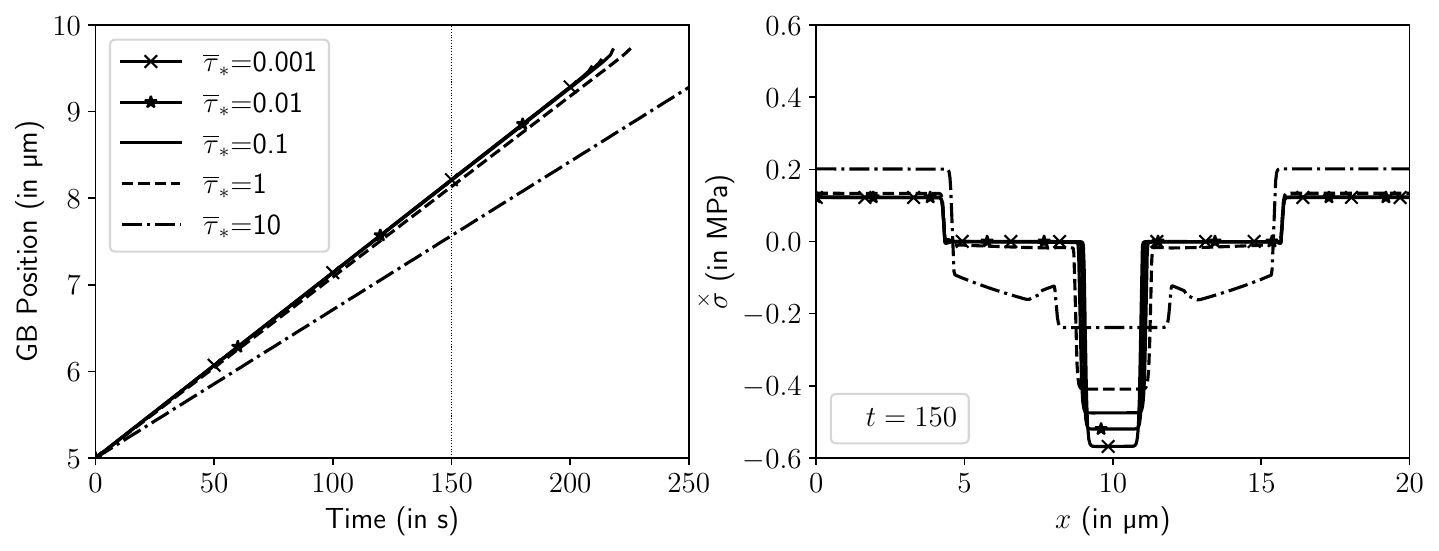}
	\caption{Grain boundary position in time for different $\tau_*$ (left), and skew-stress $\overset{\times}{\sigma}$ profile for each at $t=150$ seconds (right).}
	\label{fig:gbmigtaustar}
\end{figure}

The co-evolution of $\theta$ and $\pvec{e}{*}$ is crucial, and it may not be maintained depending on the value of inverse mobility $\tau_*$. Fig. \ref{fig:gbmigtaustar} (left) shows the grain boundary position in time for different $\overline{\tau}_*$. It seems that above a certain threshold, the motion is disturbed and the grain boundary velocity decreases. When $\overline{\tau}_*$ is small enough, the velocity is not affected. As also observed by \cite{ask2018cosserat}, a good rule of thumb is to set the mobility $\overline{\tau}_*$ smaller than phase field mobility; hence, we have used $\overline{\tau}_*=0.01$ and $\tau_\eta=0.1$ in this section. Fig. \ref{fig:gbmigtaustar} (right) shows the $\accentset{\times}{\sigma}$ at $t=150$ s for different $\overline{\tau}_*$. When $\overline{\tau}_*$ is not sufficiently small, the evolution of $\pvec{e}{*}$ lags behind $\theta$, and the skew stress is not fully relaxed in the wake of the grain boundary. In addition, the final coalescence and the annihilation of the inner grain happens slightly earlier for smaller $\overline{\tau}_*$, as evidenced by small peaks in Fig. \ref{fig:gbmigtaustar} (left) towards the end, and higher values of $\accentset{\times}{\sigma}$ on Fig. \ref{fig:gbmigtaustar} (right).

\section{Conclusion}\label{sec:conc}

We have presented a new multi-physics model coupling HMP type orientation phase field with the Cosserat crystal plasticity framework for evolution of grain boundary microstructure. The formulation of the model is inspired by the work of \cite{ask2018cosserat}, which couples Cosserat mechanics to KWC orientation phase field. The proposed model offers several improvements compared to the previous one, while keeping the same novel mechanics and capabilities. 

The main feature of the proposed model is the ability of the lattice orientation to evolve concurrently with both grain boundary migration and mechanical deformation. Firstly, the curvature driven grain boundary evolution mechanism is inherited from the HMP orientation phase field model, which was validated with a triple-junction test. Secondly, the lattice orientation is represented by the Cosserat micro-rotation through a constraint on the constitutive level. This makes its evolution by mechanical deformation consistent, which is tested with the elastic deformation of a periodic bicrystal. Similar to KWC coupled model, an eigendeformation measure is introduced so that the stresses in the undeformed state are zero.

Another driving force for grain boundary migration is the non-homogeneous distribution of SSDs, which was incorporated into KWC with an SSD energy term \citep{abrivard2012aphase}. However, this enhancement undesirably alters the equilibrium state of the order parameter. We applied a modified form to the HMP coupled model so that the same mechanism is preserved without altering the phase field equilibrium which is an improvement of the previous work.

By construction, the HMP model removes non-physical long range interactions between the grain boundaries that exist in the KWC model due to the $|\nabla\theta|$ term in the free energy. 
Moreover, the employed singular coupling function $g(\eta)$ can be modified in order to obtain the desired misorientation dependence of the grain boundary energy without changing the variational form of the model; examples are Read-Shockley type or inclination dependent anisotropic energy.


While the HMP based model is a mathematically improved formulation compared to KWC, it is still highly non-linear and thus computationally heavy. Therefore, in this work we have shown the fundamental capabilities of the proposed model with simple numerical examples. In future work, inclination dependent grain boundaries, plastic deformation induced microstructure evolution and sub-grain formation in polycrystals will be explored.

\section{Acknowledgment}
We are grateful to Dr. Anna Ask and Prof. Samuel Forest for their insight and valuable discussions.

\appendix
\section{Model parameters}\label{app:param}

At this initial stage of the model development, we are conducting qualitative tests and thus have not exactly fitted the parameters to experimental data. Nevertheless, the parameters are selected to be in an acceptable range for pure copper. For the mechanical part of the model, the elasticity constants for isotropic elasticity and cubic anisotropy are available in the literature. The Cosserat couple modulus $\mu_c$ is a penalty parameter preventing differences between Cosserat microrotation $\dot{\ubar{\Theta}}$ and lattice rotation $\pvec{\omega}{}$. Hence, it needs to be large enough, ideally some orders higher than elastic shear modulus $\mu^e$. The parameters of the orientation phase field model $\alpha$, $\nu$ and $\mu$ should be selected carefully so that the grain bulk and boundary are clearly distinguished while staying in the limitations of the model. These parameters determine the width, energy and the mobility of grain boundaries. For a grain boundary without inclination dependence the 1D equilibrium profiles of order parameter $\eta$ and orientation $\theta$ can be obtained with asymptotic analysis as outlined in \cite{henry2012orientation} and \cite{staublin2022phase}. However, we first need to non-dimensionalize our free energy functional.

A length scale $L$ with unit m is defined where $(\ol{x},\ol{y},\ol{z})=(x,y,z)/L$ are dimensionless coordinates. With $L$ we can non-dimensionalize following parameters and differential operator,

\begin{equation}
	\ol{\alpha}=\alpha,\quad \ol{\nu}=\nu/L,\quad \ol{\mu}=\mu/L,\quad \ol{\nabla}=L\nabla,\quad \ol{C}_A=C_A/L,\quad \ol{C}_*=C_*/L
\end{equation}
The mechanical parameters are non-dimensionalized using $f_0$ with unit Pa as

\begin{equation}
	\ol{\tenq{E}}^s=\tenq{E}^s/f_0,\quad \ol{\mu}_c=\mu_c/f_0,\quad \ol{\lambda}=\lambda,\quad \ol{\mu}^e=\mu^e/f_0,\quad \ol{r}^\alpha = r^\alpha.
\end{equation}
Finally, the non-dimensional free energy functional is given by

\begin{align}
	\begin{split}
		\ol{\psi}=\dfrac{\psi}{f_0}&= \left[\ol{\alpha} V(\eta)+\frac{\ol{\nu}^2}{2}|\ol{\nabla}\eta|^2+\ol{\mu}^2g(\eta)|\ol{\nabla}\theta|^2\right]\\
		&+\dfrac{1}{2}\tend{\varepsilon}^e:\tenq{\ol{E}}^s:\tend{\varepsilon}^e+2\ol{\mu}_c\,\accentset{\times}{e}^e\,^2+\phi(\eta)\sum_{\alpha=1}^{N}\frac{\ol{\lambda}}{2}\ol{\mu}^e\ol{r}^{\alpha\,2}.
	\end{split}
\end{align}
The total energy can be calculated by

\begin{equation}
	\ol{\Psi}=\dfrac{\psi}{f_0L^3}=\int_{\ol{\Omega}}\ol{\psi}d\ol{\Omega},
\end{equation}
where $d\ol{\Omega}=L^3d\Omega$. By defining a time scale $t_0$ with $\ol{t}=t/t_0$, we can non-dimensionalize mobility constants

\begin{equation}
	\ol{\tau}_\eta=\dfrac{\tau_\eta}{f_0t_0},\quad \ol{\tau}_*=\dfrac{\hat{\tau}_*}{f_0t_0}.
\end{equation}
The equilibrium profile between two semi-infinite grains with different lattice orientations can be found by considering equations \eqref{eqn:hmpetadot} and \eqref{eqn:tdot_hmpccp} without dislocations and setting rates to zero,

\begin{align}
	0&=\nu^2\dfrac{\partial^2\eta}{\partial x^2}-\alpha V_{,\eta}+\mu^2g_{,\eta}\left(\dfrac{\partial\theta}{\partial x}\right)^2\\
	0&=\dfrac{\partial}{\partial x}\cdot\left[g(\eta)\dfrac{\partial\theta}{\partial x}\right]
\end{align}
These equations are solved by dividing the space into an inner region with rapid change of variables and an outer region where change is small. Then integration constants are found by equating the inner and outer region solutions at the intersection point. The field variables in the inner region are expanded in terms of a small constant $\epsilon$,

\begin{equation}
	\begin{split}
		\eta = \eta^0+\epsilon\dfrac{\partial\eta}{\partial x}+\epsilon^2\dfrac{\partial^2\eta}{\partial x^2} + ... \\
		\theta = \theta^0+\epsilon\dfrac{\partial\theta}{\partial x}+\epsilon^2\dfrac{\partial^2\theta}{\partial x^2} + ...
	\end{split}
\end{equation}
Here, we only provide the final forms and the reader is referred to \cite{henry2012orientation} and \cite{staublin2022phase} for details. Leading order solutions are given by,

\begin{align}
	\dfrac{\partial\theta^0}{\partial z}&=\dfrac{A}{g(\eta^0)}\label{eqn:theta0} \\
	\dfrac{\partial\eta^0}{\partial z} &= \pm\dfrac{\sqrt{2}}{\ol{\nu}}\sqrt{\ol{\alpha}V(\eta^0)-\dfrac{A^2\ol{\mu}^2}{g(\eta ^0)}} \label{eqn:eta0}
\end{align}
where $z=x/\epsilon$ is stretched coordinate, $A$ is a constant and \eqref{eqn:eta0} changes sign at the middle of grain boundary. The grain boundary energy $\tilde{\gamma}_{gb}$ is given by

\begin{equation}\label{eqn:gben}
	\tilde{\gamma}_{gb}(\Delta\theta)=2\sqrt{2}\ol{\nu}\int_{\eta^0_{min}}^1\dfrac{\ol{\alpha}V(\eta^0)}{\sqrt{\ol{\alpha} V(\eta^0)-\dfrac{A^2\ol{\mu}^2}{g(\eta^0)}}}d\eta,
\end{equation}
and the true grain boundary energy is

\begin{equation}\label{eqn:gbentrue}
	\gamma_{gb}=f_0L\epsilon\,\tilde{\gamma}_{gb}.
\end{equation}
To solve the preceding equations, one can start with an $\eta^0$ slightly smaller than 1, pick a value for constant $A$ and then first solve \eqref{eqn:eta0} using finite difference. Then using $\eta^0$ one can similarly calculate $\theta^0$ with \eqref{eqn:theta0}, which gives a certain misorientation $\Delta\theta$ depending on chosen value of $A$. Finally, $\tilde{\gamma}_{gb}(\Delta\theta)$ for given misorientation can be found with \eqref{eqn:gben}.

\begin{figure}[ht]
	\centering
	\includegraphics[width=1\textwidth]{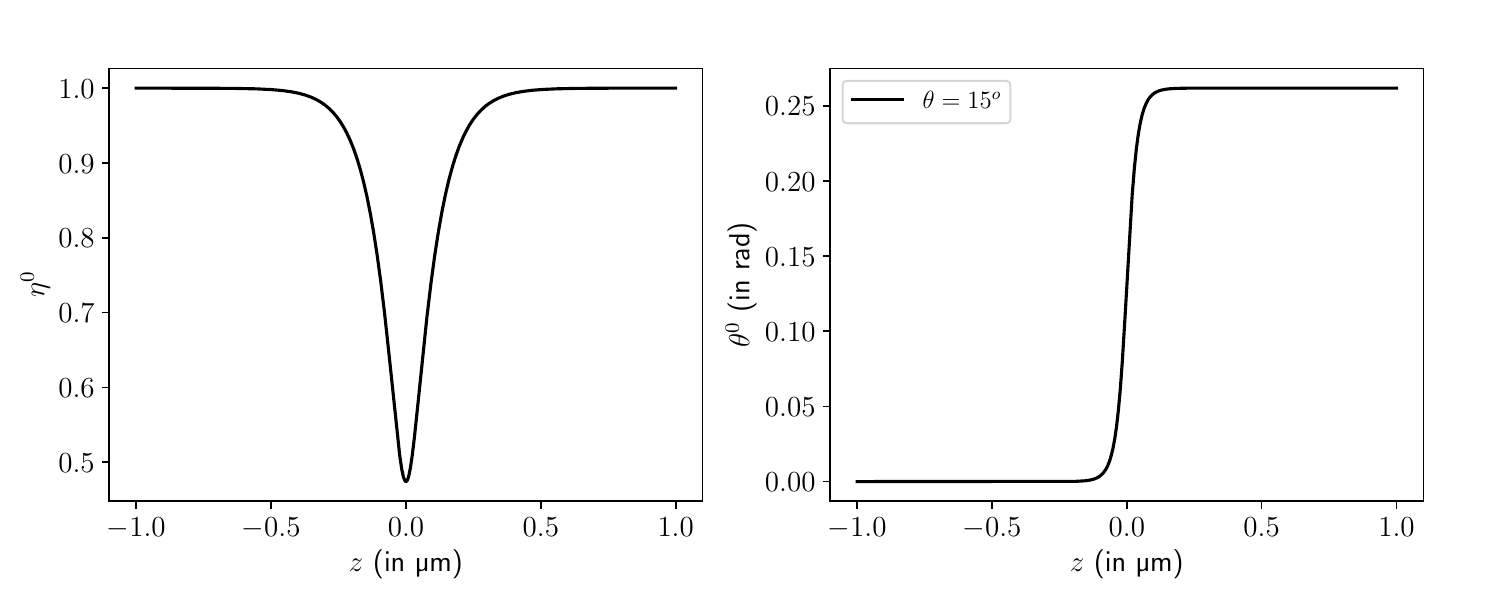}
	\caption{Equilibrium profiles of order parameter and orientation field for given misorientation found from asymptotic analysis with parameters $\ol{\alpha}=150$, $\ol{\nu}=1$, $\ol{\mu}=2.5/\pi$ and $\epsilon=1$.}
	\label{fig:asymprof}
\end{figure}

Fig. \ref{fig:asymprof} shows an example of equilibrium profiles for different misorientations obtained from asymptotic analysis with parameters $\ol{\alpha}=150$, $\ol{\nu}=1$, $\ol{\mu}=2.5/\pi$ and $\epsilon=1$. When $\ol{\nu}$ is kept constant, increasing $\ol{\alpha}$ results in a thinner grain boundary, and increasing $\ol{\mu}$ gives a deeper profile of $\eta^0$. It is important to have a sufficiently large $\ol{\mu}$ so that $\eta$ profile is not too shallow for small misorientations.

For mesoscale, an appropriate length scale is $L=1$ \textmu m. Looking at Fig. \ref{fig:asymprof}, this results in a diffuse grain boundary width of approximately 400 nm, which is huge considering a real grain boundary width, but a common practice in diffuse models to represent grain boundary with a reasonable amount of finite elements.

\begin{figure}[ht!]
	\centering
	\includegraphics[width=0.7\textwidth]{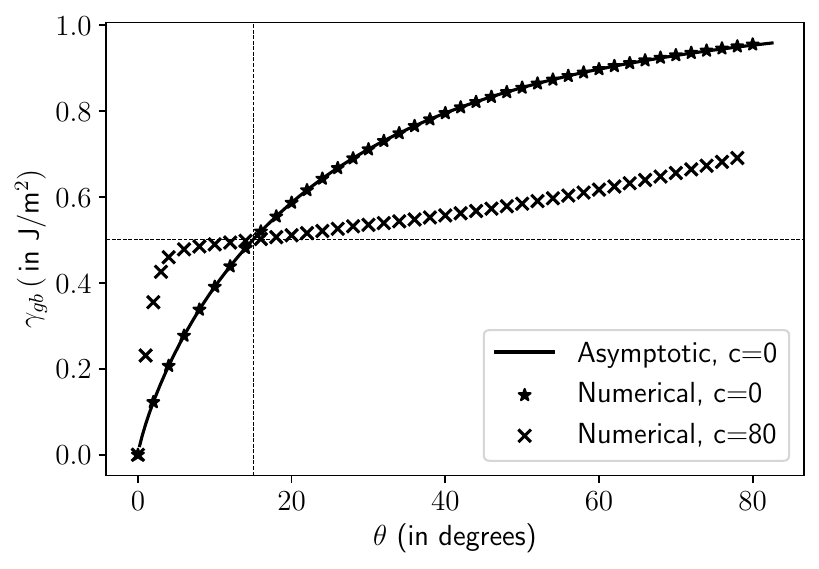}
	\caption{Grain boundary energies at increasing misorientations calibrated to an approximate 0.5 $J/m^2$ at 15$^o$ misorientation for pure Copper. Two types of curves are shown: the black curve using $g(\eta)$, the red curve where $g(\eta)$ is replaced with $g(\eta)+clog(1-\eta)+C_0$ to generate Read-Shockley type grain boundary energy. The values of $f_0$ are 87 kJ/m$^3$ and 313.6 kJ/m$^3$, respectively.}
	\label{fig:gben}
\end{figure}

Using equations \eqref{eqn:gben} and \eqref{eqn:gbentrue}, model parameters can be calibrated to fit experimental grain boundary energy data. In this work, we do not pursue a rigorous fitting. Instead, we assume an average grain boundary energy of 0.5 J/m$^2$ at 15 degrees misorientation for copper, which is in acceptable range (\cite{tschopp2015symmetric}). Using equation \eqref{eqn:gbentrue}, this gives $f_0=87$ kJ/m$^3$. The resulting grain boundary energy up to 80 degrees of misorientation is shown in Fig. \ref{fig:gben} with black curve. The markers show the energy calculated numerically with the finite element method, and there is excellent agreement with asymptotic analysis. 

A feature of HMP type orientation phase field is that if the singularity at $g(1)$ is preserved, arbitrary functions can be added to $g(\eta)$ to modify the dependency of grain boundary energy on misorientation without changing the free energy or evolution equations, for example inclination dependence (\cite{henry2012orientation}). Similarly, a Read-Shockley type grain boundary energy can be obtained by making a small modification to singular function $g(\eta)$, i.e.,

\begin{equation}
	g(\eta)=\dfrac{7\eta^3-6\eta^4}{(1-\eta)^3}+c\ln(1-\eta)+C_0 \quad\text{where}\quad C_0=min\left(\dfrac{7\eta_*^3-6\eta_*^4}{(1-\eta_*)^3}+c\ln(1-\eta_*)\right)+0.01,
\end{equation}
and $\eta_*$ is the values in range [0,1] (see \cite{staublin2022phase} for details). The resulting curve with $c=80$ is plotted in Fig. \ref{fig:gben}.

The proposed coupled Cosserat model and the pure HMP phase field have the same equilibrium condition, meaning the leading order solutions from the asymptotic analysis of HMP can be used for both models to obtain equilibrium profiles and grain boundary energy. However, the next to leading order solution of grain boundary mobility presented in \cite{staublin2022phase} is no longer valid for the coupled Cosserat model because orientation update is based on balance equation \eqref{eqn:hmpthetadot} and constraint \eqref{eqn:eeskew}, instead of relaxation dynamics. In addition, the coupled model has an additional driving force for grain boundaries due to stored dislocations. Still, as discussed in \cite{ask2020microstructure}, it is possible to estimate the grain boundary mobility by relying on simple numerical test cases.

Unless otherwise mentioned, the following set of parameters presented in Table \ref{tab:param} are used in the numerical simulations.
\begin{table}[ht]
	\caption{Parameter set used for the coupled Cosserat Crystal Plasticity - HMP orientation phase field model.}\label{tab:param}
	\centering
	\begin{tabular}{l|cccccccc}
		\multicolumn{1}{c|}{\multirow{2}{*}{Phase field}} & $f_0$         & $t_0$         & $\tau_\eta$  & $\hat{\tau}_*$       & $\nu$           & $\alpha$        & $\mu$                    &                           \\ \cline{2-9} 
		\multicolumn{1}{c|}{}                             & 87 kPa  & 1 s     & $0.1f_0t_0$    & $1f_0t_0$   & 1 \textmu m & 150             & 2.5/$\pi$ \textmu m   &                           \\ \hline\hline
		\multirow{2}{*}{Mechanics}                        & C11           & C12           & C44          & $\mu_c$        & $\lambda$       & $b$             & $C_A$                    & \multicolumn{1}{c}{$C_D$} \\ \cline{2-9} 
		& 160 GPa & 110 GPa & 75 GPa & 750 GPa & 0.3             & 0.2556 nm & $\sqrt{10}$ \textmu m & \multicolumn{1}{c}{100}  
	\end{tabular}
\end{table}

\section{Analytic solution for elastic shear loading of periodic bicrystal}\label{app:shearan}
Here a summary of the equations of analytic solution to elastic shear loading of periodic bicrystal with cubic anisotropy is presented. For the derivation and details please refer to \cite{ask2018cosserat}.

\begin{figure}[ht]
	\centering
	\includegraphics[width=1\textwidth]{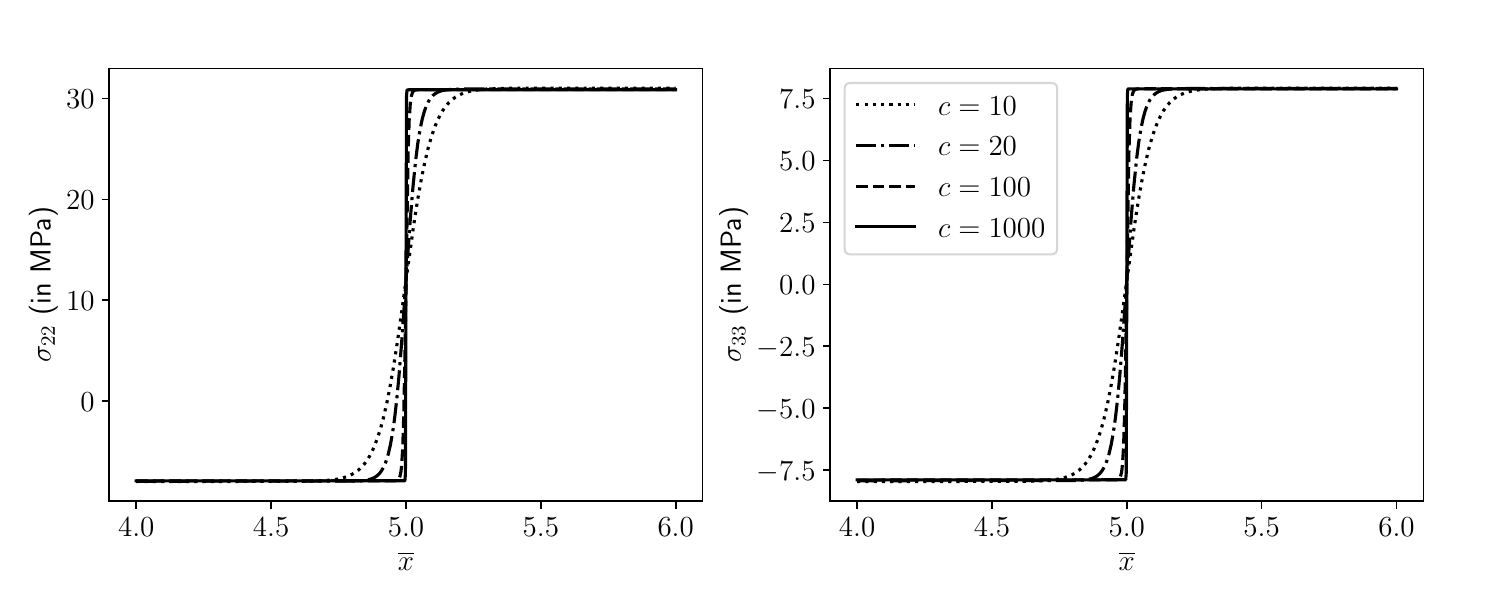}
	\caption{Stress profiles along grain boundary with changing orientation with different sharpness.}
	\label{fig:sigan}
\end{figure}

Loading is applied by imposing a mean displacement field with periodic fluctuations. The total displacement $\underline{\bm{u}}$ is given by

\begin{equation}
	\underline{\bm{u}}=\tend{B}\cdot\underline{\bm{x}}+\underline{\bm{v}},
\end{equation}
where $\underline{\bm{x}}$ is spatial coordinates, $\underline{\bm{v}}$ is the periodic fluctuation vector and $\tend{B}\cdot\underline{\bm{x}}$ is mean displacement field. The tensor $\tend{B}$ is defined as

\begin{equation}
	\tend{B}=
	\begin{bmatrix}
		0 & B_{12} & 0\;\;\\
		B_{21} & 0 & 0\\
		0 & 0 & 0
	\end{bmatrix}
\end{equation}
for shear loading. For simple shear either $B_{12}$ or $B_{21}$ is zero. Assuming $B_{12}=0$, displacements are given by,

\begin{align}
	\begin{split}
		u_1=& v_1(x_1)\\
		u_2=&B_{21}x_1+v_2(x_1)\\
		u_3=&0.
	\end{split}
\end{align}
Assuming small deformations and plane strain conditions, the only non-zero strains are

\begin{equation}\label{eqn:strains}
	\varepsilon_{11}=\dfrac{\partial v_1}{\partial x_1},\qquad \varepsilon_{12}=\varepsilon_{21}=\dfrac{1}{2}\left(B_{21}+\dfrac{\partial v_2}{\partial x_1}\right).
\end{equation}
Assuming cubic elasticity, the stiffness tensor and the compliance tensor can be defined with three independent constants each: $C_{11}$, $C_{12}$ and $C_{44}$, or $S_{11}$, $S_{12}$ and $S_{44}$, respectively. We have the following relations between them,

\begin{equation}
	S_{11}=\dfrac{C_{11}+C_{12}}{(C_{11}+2C_{12})(C_{11}-C_{12})},\qquad S_{12}=\dfrac{-C_{12}}{(C_{11}+2C_{12})(C_{11}-C_{12})},\qquad S_{44}=\dfrac{1}{C_{44}}.
\end{equation}
Assume we have a varying orientation field $\theta$ along $x_1$ direction. The constitutive relation is applied at the material frame and then $\theta$ is used to obtain strains in the global frame in terms of stresses. Then enforcing constraints \eqref{eqn:strains} results in a system of equations with unknown stresses $\sigma_{12}$, $\sigma_{11}$ and $\sigma_{22}$. After solving the system, $\sigma_{33}$ is found from plane strain conditions. We get

\begin{equation}
	\sigma_{12}=\dfrac{1}{2F}B_{21},\qquad \sigma_{11}=E\sigma_{12},\qquad \sigma_{22}=-\dfrac{B}{A}\sigma_{11}-\dfrac{C}{A}\sigma_{12},\qquad \sigma_{33}=-\dfrac{S_{12}}{S_{11}}(\sigma_{11}+\sigma_{22}),
\end{equation}
where the orientation dependent coefficients are given by

\begin{equation}
	E=\dfrac{\left<C/A\right>}{1-\left<B/A\right>},\qquad F=\left<-\dfrac{C+CE}{2}-\dfrac{CEB+C^2}{2A}+D\right>,
\end{equation}
where $<\cdot>$ represent average and

\begin{equation}
	\begin{split}
		A=S_{11}+S_D+S_A\sin^2(2\theta),\quad B=S_{12}+S_D-S_A\sin^2(2\theta),\hspace{1.5cm}\\ 
		C=S_A\sin(4\theta),\quad
		D=S_B+S_A\cos(4\theta),\hspace{3cm}\\
		S_A=\dfrac{1}{2}\left(-S_{11}+S_{12}+\dfrac{1}{2}S_{44}\right),\quad S_B=\dfrac{1}{2}\left(S_{11}-S_{12}+\dfrac{1}{2}S_{44}\right),\quad S_D=-\dfrac{S_{12}^2}{S_{11}}.
	\end{split}
\end{equation}
An orientation field in the form

\begin{equation}\label{eqn:thetaan}
	\theta = \dfrac{\pi}{24}\left[\tanh(c[\overline{x}+\overline{x}_{shift}])+1\right],
\end{equation}
gives a change from 0 to 15 degrees around position $\overline{x}_{shift}$ with different sharpness determined by $c$. Fig. \ref{fig:sigan} shows the stresses obtained using elasticity parameters in Table \ref{tab:param} and equation \eqref{eqn:thetaan}.

\section{2D numerical implementation in FEniCS}\label{app:fenics}

The model is implemented in the FEniCS 2019 open-source finite element library (\cite{logg2012automated}, \cite{alnaes2015fenics}). After the simplifications in Section \ref{ssec:freeen}, for a single slip system, the strong form equation are given by

\begin{align}
	\tau_\eta\dot{\eta}&=f_0\nu^2\nabla^2\eta-f_0\left[\alpha V_{,\eta}+\mu^2g_{,\eta}|\nabla\theta|^2\right]-\phi_{,\eta}\psi_{\rho} \\
	0&=f_0\nabla\cdot\left[\mu^2g(\eta)\nabla\theta\right]+\accentset{\times}{\sigma} \\
	\ubar{0}&=\nabla\cdot\tend{\sigma},
\end{align}
where

\begin{equation}
	V(\eta)=\dfrac{1}{2}(1-\eta)^2,\qquad g(\eta)=\dfrac{7\eta^3-6\eta^4}{(1-\eta)^3},\qquad \phi(\eta)=-2\eta^4+3\eta^3-\eta^2+\eta,\qquad \psi_{\rho}=\frac{\lambda}{2}\mu^eb^2\rho.
\end{equation}
The weak form is obtained by multiplying the strong forms with test functions $w_\eta$, $w_\theta$, $\ubar{w}_u$ and integrating. We get,

\begin{align}
	&\int_\Omega \left\{f_0\nu^2\nabla\eta\cdot\nabla w_\eta + \left(\tau_\eta\dot{\eta}+f_0\left[\alpha V_{,\eta}+\mu^2g_{,\eta}\mathcolorbox{lightgray}{|\nabla\theta|}^2\right]+\phi_{,\eta}\psi_{\rho}\right)w_\eta\right\} dV = \int_{\partial \Omega}f_0\nu^2\nabla\eta\cdot\ubar{n}\;w_\eta dS \label{eqn:fenicsstrongeta}\\
	&\int_\Omega \left\{f_0\mu^2g(\eta)\nabla\theta\cdot\nabla w_\theta -\accentset{\times}{\sigma}w_\theta\right\} dV = \int_{\partial \Omega}f_0\mu^2g(\eta)\nabla\theta\cdot\ubar{n} \;w_\theta dS \label{eqn:fenicsstrongtheta}\\
	&\int_\Omega \left\{\tend{\sigma}^s(\ubar{u}) : \tend{\varepsilon}(\ubar{w}_u)+2\pvec{\sigma}{}(\ubar{u})\cdot\pvec{\omega}{}(\ubar{w}_u)\right\} = \int_{\partial\Omega} (\tend{\sigma}\cdot\ubar{n})\cdot \ubar{w}_u dS
\end{align}
The equations are discretized in time implicitly, except for the highlighted term in \eqref{eqn:fenicsstrongeta} where the value in the beginning of the increment is used. The state variables are updated with the following equations,

\begin{equation}
	\Delta \pvec{e}{*}=\dfrac{\dfrac{\Delta t2\mu_c}{\hat{\tau}_*g(\eta)}(\pvec{e}{}-\pvec{e}{*}_{n}-\pvec{e}{p})}{1+\dfrac{\Delta t2\mu_c}{\hat{\tau}_*g(\eta)}}
\end{equation}

\begin{equation}
	\Delta \rho_{\text{recovery}}=
	\begin{cases}
		-\rho C_D\tanh(C_A^2|\nabla\theta_n|^2)\Delta\eta \quad&\text{if}\quad \Delta\eta>0,\\
		0 &\text{if}\quad \Delta\eta\le 0,
	\end{cases}
\end{equation}
where subscript $n$ represents the value in the beginning of the increment. 

The system of nonlinear equations are solved monolithically with Newton-Raphson iteration. During the iterations, due to the singular nature of $g(\eta)$, the residual from equation \eqref{eqn:fenicsstrongtheta} is significantly bigger relative to the residual from \eqref{eqn:fenicsstrongeta}. Therefore, in order to solve these equations monolithically, the former is normalized with a coefficient of $10^{-10}$. Moreover, to increase the numerical stability, the singular function $g(\eta)$ and its derivative $g_{,\eta} (\eta)$ is modified with a cut-off value as follows,

\begin{equation}
	g(\eta) \rightarrow g(min(\eta,\eta_{\,\text{cutoff}}))
\end{equation}
where $\eta_{\,\text{cutoff}}$ was taken as $(1-10^{-4})$ or $(1-10^{-5})$.

\bibliographystyle{elsarticle-harv} 
\bibliography{mybibfile}

\end{document}